\documentclass[12pt]{article}
\usepackage{amsmath,amssymb,url,mathrsfs, nicefrac, color}
\usepackage[semicolon]{natbib}
\usepackage[capposition=bottom]{floatrow}
\usepackage{mwe}
\usepackage[skins]{tcolorbox}
\usepackage{lipsum}
\usepackage{comment}
\definecolor{myred}{RGB}{213,94,0}
\definecolor{mygreen}{RGB}{0,158,115}
\definecolor{myblue}{rgb}{0,0,0.75}
\definecolor{bcblue}{RGB}{0,30,52}

\usepackage{hyperref}
\hypersetup{
  colorlinks   = true,    
	urlcolor     = myblue,    
	  linkcolor    = myblue,    
		citecolor    = myblue, 
		 breaklinks=true,   
           colorlinks=true,   
	}
\usepackage{tikz}
\usetikzlibrary{patterns, calc,shapes,decorations.pathreplacing,positioning, shapes.multipart}
\usepackage{pgfplots}
\usepackage{subfig}
\usepgfplotslibrary{groupplots}
\usetikzlibrary{arrows.meta} 
\usepackage{amsthm}
\makeatletter
\def\th@plain{%
\thm@notefont{}
  \itshape 
}
\def\th@definition{%
  \thm@notefont{}
	\normalfont 
}
\makeatother
\usepackage{mathtools}
\usepackage[english]{babel}
\usepackage{color}
\usepackage{bm}
\usepackage{accents}	
\usepackage{graphicx}
\usepackage[semicolon]{natbib}
\theoremstyle{plain}

\newtheorem{lemma}{Lemma}
\newtheorem{proposition}{Proposition}
\newtheorem{corollary}{Corollary}

\theoremstyle{definition}
\newtheorem{definition}{Definition}
\newtheorem*{definition*}{Definition}
\newtheorem*{assumption*}{Assumption}
\newtheorem{assumption}{Assumption}
\newtheorem*{conjecture*}{Conjecture}

\usepackage{url}
\usepackage{booktabs}

\usepackage[top=1in, bottom=1in, left=1in, right=1in]{geometry}
\usepackage[onehalfspacing]{setspace}
\newcommand{\reals}{\ensuremath{\mathbb{R}}}
\newcommand{\expect}{\ensuremath{\mathbb{E}}}

\newcommand{\lb}{\ensuremath{\left(}}
\newcommand{\rb}{\ensuremath{\right)}}

\newcommand{\w}{\ensuremath{w}}
\newcommand{\wone}{\ensuremath{w_1}}
\newcommand{\wtwo}{\ensuremath{w_2}}
\newcommand{\wL}{\ensuremath{w_L}}
\newcommand{\wH}{\ensuremath{w_H}}

\newcommand{\ts}{\ensuremath{\tilde{s}}}
\newcommand{\MR}{\ensuremath{\varphi}}
\newcommand{\MRI}{\ensuremath{\bar\MR}}

\newcommand{\qh}{\ensuremath{\hat{q}}}

\newcommand{\sigmah}{\ensuremath{\hat{\sigma}}}
\newcommand{\ind}{\ensuremath{r}}

\pgfplotsset{compat=1.17} 
 \title{Menu Pricing of Large Language Models\thanks{Bergemann: Department of Economics, Yale University, dirk.bergemann@yale.edu. Bonatti: MIT Sloan, bonatti@mit.edu. Smolin: Toulouse School of Economics, alexey.v.smolin@gmail.com. We thank Mark Armstrong, Mert Demirer, Scott Kominers, Antonio Russo, Ron Siegel, and Frank Yang for valuable comments and discussions, as well as audiences at the Triangle Conference, USC, Caltech, Virginia, the Chicago workshop, the Sciences Po workshop, the NBER Market Design workshop, VSET, Brown, LSE, the TSE Digital Economics Conference, the IESE Economics of AI Conference, and Cambridge. An extended abstract of an earlier version of this paper appeared as ``The Economics of Large Language Models: Token Allocation, Fine-Tuning, and Optimal Pricing'' in the proceedings of EC'25. Dirk Bergemann gratefully acknowledges financial support from NSF SES 2049754 and ONR MURI. Alessandro Bonatti gratefully acknowledges financial support from NSF SES 2519401. Alex Smolin gratefully acknowledges funding from the French National Research Agency (ANR) under the Investments for the Future program
(grant ANR-17-EURE-0010) and through the AI Interdisciplinary Institute ANITI (grant ANR-23-IACL-0002).}}
    \author{Dirk Bergemann \and Alessandro Bonatti\and Alex Smolin}
    \date{\today}

\begin{document}
    \maketitle
\thispagestyle{empty}

\begin{abstract}
We develop a framework for the optimal pricing and product design of LLMs in which a provider sells menus of token budgets to users who differ in their valuations across a continuum of tasks. Under a homogeneous production technology, we show that users' high-dimensional type profiles are summarized by a scalar index, reducing the seller's problem to one-dimensional screening. The optimal mechanism takes the form of committed-spend contracts: buyers pay for a budget that they allocate across token classes priced at marginal cost. We extend the analysis to environments with multiple differentiated models and to competition between a proprietary leader and an open-source fringe, showing that competitive pressure reshapes both the intensive and extensive margins of compute provision. Each element of our theory (token-budget menus, maximum- and minimum-spend plans, multi-model versioning, and linear API pricing) has a direct counterpart in the observed pricing practices of providers such as Anthropic, OpenAI, and GitHub.\bigskip

\end{abstract}

\newpage
\setcounter{page}{1}
\section{Introduction}

Access to large language models is fast becoming a major input to economic activity. Businesses use LLMs for coding, customer support, legal research, and content generation; individual users rely on them for writing, analysis, and decision-making. The leading providers, Anthropic, OpenAI, and Google, collectively serve hundreds of millions of users and generate billions of dollars in annualized revenue. Yet the pricing of these services remains strikingly ad hoc: subscription tiers, token-based metering, credit systems, and volume commitments coexist with no clear theoretical foundation. Understanding how a profit-maximizing provider should price LLM access, and quantifying the distortions that arise when it does, is the goal of this paper.\footnote{For the practical challenges associated with LLM pricing, see \url{https://www.wsj.com/articles/no-one-knows-how-to-price-ai-tools-f346ea8a}.}

Pricing access to LLMs is  fundamentally a multidimensional screening problem. An LLM user faces a continuum of tasks, each with a different value. The provider offers token budgets---bundles of input, output, and fine-tuning tokens---that the user allocates across tasks at her discretion. The provider can meter total token usage but cannot observe or contract on the user's task-by-task allocation, creating a combined adverse-selection and moral-hazard problem. User types are infinite-dimensional (a function from tasks to values), the allocation space is high-dimensional (tokens of multiple classes across many tasks), and the user's hidden action (token allocation across tasks) further compounds the difficulty. A priori, this problem seems intractable.

Our central result is that it is not. Under a homogeneous production technology---meaning that the gain function is homogeneous, so the optimal composition of inference tokens is scale-invariant---each user's entire type profile collapses to a scalar index, the \emph{aggregate type}. Users with the same aggregate type make the same total token demands, receive the same total surplus, and choose the same menu item, regardless of the fine structure of their task valuations. This aggregation property reduces the provider's problem to one-dimensional screening \`{a} la \cite{mussa1978monopoly}, despite the underlying complexity of the environment. Building on this reduction, we develop the analysis in four steps.

First, we characterize the efficient allocation (\autoref{sec:efficient}). Efficiency requires that all tasks employ inference token classes in common proportions; only the scale of token usage varies with a task's marginal value. When the provider faces capacity constraints in each token class, the efficient allocation can be implemented through linear prices equal to inflated shadow costs (\autoref{cor:capacity-efficient}), a result that rationalizes the linear, pay-per-token pricing universally observed in developer-facing API markets.

Second, we characterize the optimal mechanism for a single-model monopolist (\autoref{sec:packages}). By the aggregation result, the buyer's indirect utility from a token-budget bundle takes a multiplicative form: aggregate type times aggregate quality. The optimal menu therefore excludes low types and distorts quality downward for (almost) all others, exactly as in the standard one-dimensional framework. Crucially, the optimal mechanism admits intuitive indirect implementations (\autoref{prop:indirect-implementation}): it can be realized as a maximum-spend mechanism (a budget of credits allocated across tokens priced at marginal cost), as a minimum-spend mechanism (volume commitments that unlock lower per-token prices), or as a menu of two-part tariffs. These implementations are not merely theoretical curiosities; they correspond precisely to the pricing structures observed at leading platforms.

Third, we extend the framework to multiple differentiated models (\autoref{sec:multi-model}). When models share a common returns-to-scale parameter in inference but differ in returns to fine-tuning, the buyer's payoff has a constant elasticity of substitution over model qualities. Cost minimization implies that each type uses a single model for all tasks (\autoref{prop:packages-multiple}). This generates a natural versioning structure: higher-tier plans grant access to more capable models, not merely larger usage allocations. Anthropic's pricing illustrates the single-model benchmark of Section~\ref{sec:packages}: all paid tiers access the same models, with differentiation occurring through compute budgets. OpenAI, by contrast, screens jointly on usage and model access, reserving its most compute-intensive reasoning model for the highest tier, consistent with the multi-model menu of \autoref{sec:multi-model}.

Fourth, we study competition between a proprietary leader and an open-source fringe that sells tokens at marginal cost (\autoref{sec:multi-model}). The leader designs an optimal menu subject to the buyer's ability to supplement usage from the fringe. Three regions emerge: low types purchase exclusively from the fringe; intermediate types buy from the leader at quantities that exactly deter fringe top-up; and high types are served as under pure monopoly, with standard downward distortions (\autoref{prop:fringe}). Competition thus reshapes both the intensive margin (how much compute is sold to each buyer) and the extensive margin (which buyers adopt the proprietary model at all).

We conclude by connecting each theoretical construct to observed pricing (\autoref{sec:application}). Consumer subscriptions at Anthropic and OpenAI implement the nonlinear menus of Sections~\ref{sec:packages} and~\ref{sec:multi-model}. Model aggregators (e.g., Quora's Poe and GitHub Copilot) implement the committed-spend mechanisms of \autoref{sec:two-part-tariff}, differing in whether they enforce a hard budget cap (maximum spend) or allow overages at linear prices (minimum spend). API pricing is  linear in tokens with no volume discounts, consistent with providers prioritizing adoption over rent extraction. Thus, pricing practices observed in the industry  suggest that these mechanisms capture fundamental economic forces rather than incidental design choices.

\paragraph{Related Literature} 
Our paper contributes to the rapidly growing literature on the economics of AI. Existing work has examined the impact of LLMs on price competition \citep*{fish2024algorithmic}, token auctions \citep*{duetting2024mechanism}, and sponsored search \citep*{bergemann2024data}, and \cite*{defr25} document pricing and market-share patterns. By contrast, our focus is on the provider's own design problem, how to price and version LLM access, which has received surprisingly little theoretical attention. For example, \cite*{mahmood2024pricing} studies competition among producers of generative AI who can specialize in different tasks, but restricts attention to linear token pricing. We go further and characterize the fully optimal nonlinear mechanism, leveraging the sufficient-statistic reduction generated by the homogeneity of the gain function, which renders the joint screening-and-moral-hazard problem solvable and yields sharp, implementable predictions.

At their core, LLMs are an information technology, connecting our work to the literature on selling information \citep*{babaioff2012optimal, bergemann2018design, yang2022selling}. However, LLMs possess distinctive features: they are general-purpose technologies deployed across many tasks, usage is metered in tokens, and precision is improved by combining inference and fine-tuning, which give rise to a structurally different design problem. 

Methodologically, our aggregation result relates to the literature on ``1.5-dimensional'' mechanism design \citep*{fiat2016fedex,devanur2020optimal}. 
Unlike most of that literature, however, we must address potential incentive clashes across the continuum of subproblems that arise from the buyer's hidden allocation of tokens, requiring an additional argument based on the production technology. Even in simpler settings, such as one-dimensional screening with moral hazard \citep*{castro2024disentangling} or multidimensional screening and bundling without moral hazard \citep*{arms96,rochet2003economics,daskalakis2017strong}, tractability issues arise. Our environment combines both, and our  solution exhibits a constrained efficiency property as in \cite*{lati90} and \cite*{dodkt25}.

\section{Model}\label{sec:model}

\subsection{Task Environment}\label{subsec:task}

A buyer of LLM services  faces a unit measure of tasks indexed by $i\in[0,1]$. In our baseline setting, an  LLM provider offers one type of model only. 
The buyer can decide on (i) how many \emph{inference} tokens to use for each task $i$, $x_i\in \reals^J_+$  and (ii) how many \emph{fine-tuning} tokens to use to improve the model's performance (e.g., precision) on all tasks, $z\in \reals^K_+$. In language models, a token is a unit of text, typically a word.

The model's performance on  task $i$ is given by the \emph{gain function} $g(x_i,z)$. The function $g$ is common across tasks and takes the following form:
\begin{equation}\label{eq:gain-function}
 g(x_i,z)=\Psi(x_i)\,\Phi(z).   
\end{equation}
Thus, the gain function is multiplicatively separable in inference and fine-tuning tokens. 

Diminishing returns to inference tokens on a single task are a fundamental feature of LLM performance, consistent with empirically observed scaling laws. Thus the function $\Psi$ is positive, increasing, strictly concave, differentiable, Inada at 0,\footnote{That is, for all $j$ and $x_{i,-j}\in\reals_{+}^{J-1}$ such that $\Psi (0,x_{i,-j})>0$, $\lim_{x_{ij}\downarrow 0}\Psi_j (x_{ij},x_{i,-j})=+\infty$.} and homogeneous of degree $\sigma\in(0,1)$, i.e., $\Psi(\ind x_i)=\ind^\sigma \Psi(x_i)$ for all $\ind>0$. This function  captures the returns to inference tokens. A canonical example of such a function is a CES function, $\Psi(x_i)=\big(\sum_{j=1}^{J}\alpha_j x_{ij}^{\rho}\big)^{\sigma/\rho}$, where $\alpha_j>0, \sum_{j}\alpha_j=1,\rho<1$. The function $\Phi(z)$ is  positive, increasing, such that $g$ is strictly concave (e.g., $\Phi(z)^{1/(1-\sigma)}$ is strictly concave) and $(1-\sigma)$-Inada at 0.\footnote{\label{foot:inada}That is, there exists $z_0\geq 0$, $z_0\neq 0$, such that $\lim_{\ind\downarrow 0} \Phi(\ind z_0)/\ind^{1-\sigma}=+\infty$. Sufficient conditions for this property are (i) $\Phi(0)>0$ or (ii) $\Phi$ is homogeneous of degree $\hat\sigma\in(0,1)$ with $\sigma+\hat\sigma<1$.} This  function captures the returns to fine-tuning. 

These properties are similar in shape to the scaling laws for training LLMs. Concavity captures the fundamental assumption that time and computing resources spent on a single task exhibit diminishing marginal returns. Homogeneity is the key tractability assumption. It ensures that the optimal mix of token classes is the same for every task; only the scale of token usage varies. This property drives the aggregation result in Section~\ref{sec:efficient}, which reduces the seller's infinite-dimensional screening problem to a one-dimensional problem. The parameter $\sigma$ captures the returns to scale.

The buyer's marginal value of performance on each task is captured by $\w=(w_i)_{i\in[0,1]}$,
\begin{equation}\w:[0,1]\to[0,1],
\end{equation}
which we refer to as the buyer's \emph{type}. Using a $z$-fine-tuned model with a profile of $(x_i)_{i\in[0,1]}$ inference tokens delivers the following total payoff for buyer type $\w$:
\begin{equation}
 \int_0^1 w_i g(x_i,z) di.
\end{equation}
The buyer's type is distributed according to a commonly known  distribution  $F_{\w}$. The buyer knows his type, while the provider does not.

The provider bears the cost of processing tokens. We assume that the marginal processing costs are constant but can vary across different token classes. The cost of inference tokens is $c_j > 0$, $j\in[J]$, and the cost of fine-tuning tokens is $\hat c_k > 0$, $k\in[K]$. We let  $c\in\reals^J_{++}$ and $\hat c\in\reals^K_{++}$ denote the corresponding cost profiles. 

\paragraph{Notation} Throughout the text, we use the following notation. ``$\triangleq$'' indicates a definition.  $[K]\triangleq\{1,\dots,K\}$. $\reals_+^K\triangleq\{x\in\reals^K: x_k\geq 0,\ k\in[K]\}$ and $\reals_{++}^K\triangleq\{x\in\reals^K: x_k>0,\ k\in[K]\}$.  All stand-alone qualifiers such as ``positive,'' ``increasing,'' and ``concave'' are understood in the weak sense, that is, as ``non-negative,'' ``non-decreasing,'' and ``weakly concave,'' respectively. For constants, subscripts refer to  variables or tasks; for functions, they indicate partial derivatives. All proofs are in \autoref{app:proofs}.

\subsection{Mechanism Design}
The provider contracts on the total number of tokens of each class used by the buyer. In other words, the provider sells budgets of various kinds of tokens.  A mechanism consists of an arbitrary menu of the form
\begin{equation}
    \{(X_1(\w),\dots,X_J(\w),Z_1(\w),\dots,Z_K(\w), t(\w))\}_{\w},
\end{equation}
where $X_j$ is the total number of inference tokens of class $j$. Upon purchase, the buyer can freely  distribute those tokens across  tasks. (We relax this restriction in  \autoref{app:alloc}, where we allow the provider to contract on the usage of all tokens by the buyer.) $Z_k$ is the number of fine-tuning tokens of class $k$, which is used for fine-tuning and cannot be distributed.

The seller's problem is an optimal mechanism design problem with infinite-dimensional private information and moral hazard.  However, we show that the homogeneity of the gain function  renders the problem tractable.

\subsection{Mapping the Setting to Practice}\label{sec:mapping}

Before turning to the analysis, we describe how the primitives of our setting relate to the design and operation of contemporary large language models (LLMs).

\paragraph*{Fine-tuning tokens, \boldmath $z\in\reals^K_+$} The case $z=0$ corresponds to a baseline model that has not been fine-tuned.\footnote{For tractability, we will often study the case $g(x,0)=0$. This can be viewed as a normalization that doesn't affect the economics of the problem.} In this case, the model can be interpreted as a \emph{foundation model}. Its quality depends on its size (i.e., the number of parameters)\footnote{The parameter counts of leading models are estimated to be in the several-billion range; see \url{https://codingscape.com/blog/most-powerful-llms-large-language-models}.}, its architecture, the quantity and quality of training data, and the details of the training procedure. Scaling up model size typically improves capability and accuracy, but state-of-the-art performance generally requires commensurate scaling of training data as well \citep{kaplan2020scaling}. In our setting, we take pretraining as given. Accordingly, we focus on inference-time compute rather than training-time compute.

When $z>0$, this variable captures the number of tokens used to fine-tune the foundation model. Fine-tuning resembles pretraining in that it updates the model's parameters, without changing its size or architecture, but it is much smaller in scale. It is performed on a dataset directly relevant to a class of tasks (e.g., labeled X-ray scans, code examples, or conversational transcripts) and can be interpreted as injecting task-specific knowledge and behavior into the model. 

\paragraph*{Inference tokens, \boldmath $x_i\in\reals^J_+$} These are  the tokens used directly to process a given task. The two standard classes of inference tokens are input and output tokens. Input tokens are those provided by the user. Increasing the number of input tokens can improve predictive quality for two reasons. First, richer prompts provide more context, enabling more tailored and appropriate responses. Second, additional input may supply the data on which the model is meant to operate. For example, retrieval-augmented generation (RAG) uses the prompt to retrieve relevant passages from an external corpus and appends them to the prompt.

Output tokens are those generated by the model. A larger number of output tokens can also improve quality. First, it allows the model to provide more detailed and nuanced answers. Second, it provides opportunities for multi-step reasoning, often described as chain-of-thought (CoT) computation. This additional reasoning channel can enable the model to solve more complex tasks, at the cost of generating many intermediate tokens, often hidden from the user. This inference-time reasoning is qualitatively distinct from pretraining, remains only partially understood, and is an active frontier for improving LLM performance.\footnote{Recent progress has leveraged this channel; see, for instance, \url{https://arcprize.org/blog/oai-o3-pub-breakthrough}.}

\paragraph{Gain function, \boldmath $g$} Tokens of different classes enter the large language model as distinct inputs. They are neither perfect substitutes nor perfect complements, but each can contribute to higher predictive quality. This motivates our gain-function formulation in (\ref{eq:gain-function}), which is also in line with the empirically observed scaling laws for inference-time computation (e.g., \citealp*{wu2025inference}).

Furthermore, one can envision other contractible token categories, either by refining the distinctions above (e.g., separating reasoning tokens from response tokens) or by introducing new modalities and resources (e.g., media files, external databases, etc.). Accordingly, our model accommodates an arbitrary number of inference and fine-tuning token classes. 

\paragraph*{Costs, \boldmath $c_j,\hat c_k$} Processing tokens is costly because it requires executing the neural network and, for fine-tuning, computing parameter updates, which in turn requires energy
and specialized hardware. 
Because tokens within a given class are processed symmetrically from a computational perspective, we assume that the marginal cost per token within a class is constant across tasks.\footnote{Note that fine-tuning typically does not change the model's architecture or size and therefore should not directly affect marginal inference costs.} However, because different token classes correspond to different computational routines, we allow marginal costs to differ across classes.

\section{Efficient Solution}\label{sec:efficient}
In this section, we begin by analyzing the socially  efficient allocation of tokens to buyers and tasks. This solution provides a useful benchmark and also coincides with the optimal monopoly solution when the buyer's type is known. We then turn to the problem of a social planner who faces capacity constraints for each token class, and we use the constrained-efficient solution to characterize the buyer's  utility from purchasing a fixed token budget. 

\paragraph{Unconstrained problem} Given a type $\w=(w_i)_{i\in[0,1]}$, the efficient allocation solves the following problem: 
\begin{equation}\label{eq:planner-problem}
 \max_{(x_i)_{i\in[0,1]},z\geq0}\int_{0}^{1} w_i \Psi(x_i)\Phi(z)di-\sum_{j=1}^{J}c_j\int_{0}^{1}x_{ij}di-\sum_{k=1}^{K}\hat c_k z_k.
\end{equation}

Consider the allocation of inference tokens  first. A key implication of the homogeneity of  the production function $\Psi$ is that inference tokens are optimally employed in the same proportions: the efficient allocation of tokens across tasks differs solely by the scale at which these inputs are employed. To see this, consider 
the system of $J$ first-order conditions for each $x_i=(x_{i1},...,x_{iJ})$:
\begin{equation}\label{eq:eff-foc-text}
  w_i  \nabla\Psi (x_i)\Phi(z)=c. 
\end{equation}
By equation \eqref{eq:eff-foc-text}, for all $i$, $\nabla\Psi(x_i)$ belongs to a ray with a direction $c$. Because $\Psi$ is homogeneous and strictly concave, the ray in the space of gradients corresponds to a ray in the space of tokens. Thus, for each $i$, any optimal $x_i$ can be written as:
\begin{equation*}
    x_i= r_i d,
\end{equation*}
where $d\in\reals^J_+$ is the unique vector that solves $\nabla\Psi(d)=c$, i.e., the cost-minimizing input shares. Furthermore, because $\Psi$ is homogeneous of degree $\sigma$, $\Psi_j$ is homogeneous of degree $\sigma-1$, so that the optimal  scale is given by
\begin{equation*}
      r_i=w_i^{\frac{1}{1-\sigma}}\Phi(z)^{\frac{1}{1-\sigma}}.
\end{equation*}
Substituting back in the objective function, both the total surplus and the total consumption of each token class depend on the buyer's type $\w$ only through $\int_0^1 w_i^{1/(1-\sigma)}di$. It is then convenient to define the aggregate type $\theta(\w)$ as
\begin{equation}\label{eq:CES_type}
    \theta(\w)\triangleq \bigg(\int_0^1 w_i^{\frac{1}{1-\sigma}} di\bigg)^{1-\sigma}.
\end{equation}

Our first result establishes that the total surplus and the total amount of inference tokens depend only on the aggregate type and not on the finer details of the type profile $\w$. 

\begin{proposition}[Efficient Allocation]\label{prop:efficient}
Under the efficient allocation, all buyer types $\w$ with the same aggregate type $\theta(\w)$ consume the same number of fine-tuning tokens, consume the same total number  of inference tokens in each class, and obtain the same total payoff.  The number of inference tokens allocated to task $i$ is proportional to $w_i^{\frac{1}{1-\sigma}}$. 
\end{proposition}

\autoref{prop:efficient} has a key implication: two buyers with very different task profiles---one who values a few tasks intensely and many tasks little, and another who values all tasks moderately---consume the same total resources and obtain the same total surplus, provided they share the same aggregate type. The seller therefore cannot distinguish between them on the basis of total token consumption, nor would it want to. This indistinguishability is not an assumption but a consequence of the technology: homogeneity of $\Psi$ ensures that the efficient mix of token classes is task-independent, so only the scale of usage varies, and the aggregator~\eqref{eq:CES_type} is the natural summary of how much total scale a buyer demands.

We now show that a similar logic can be used to characterize the socially efficient solution under capacity constraints in each class of tokens (training, computation, input, output).

\paragraph{Constrained problem} We now introduce capacity constraints on each token class. This detour is not merely a generalization: the special case in which all constraints bind is precisely the problem faced by a buyer who has purchased a fixed bundle of token budgets. Thus, the constrained-efficient allocation provides the foundation for the seller's mechanism design problem in \autoref{sec:packages}.

Fix a type $\w$ and capacity constraints $X_j>0$, $Z_k>0$ for all $j,k$. The capacity-constrained planner's  problem is
\begin{align*} 
 & \max_{(x_i)_{i\in[0,1]},z\geq0}\int_{0}^{1} w_i  \Psi(x_i)\Phi(z)\,di-\sum_{j=1}^{J}c_j\int_{0}^{1}x_{ij}di-\sum_{k=1}^{K}\hat c_k z_k,\\
 & \text{s.t. } \int_{0}^{1}x_{ij}di\leq X_j,\  z_k\leq Z_k \text{ for }  j\in[J], k\in[K].
\end{align*}

Our next result establishes that the capacity-constrained efficient allocation coincides with the (unconstrained) efficient allocation for marginal costs inflated by the shadow costs of the capacity constraints, $(c',\hat c')\geq (c,\hat c)$. This observation yields an immediate implementation of the efficient solution.

\begin{corollary}[Constrained Efficient Allocation]\label{cor:capacity-efficient}\strut
\begin{enumerate}
    \item In the capacity-constrained efficient allocation, all buyer types $\w$ with the same aggregate type $\theta(\w)$ consume the same total number of fine-tuning tokens and inference tokens in each class, and they obtain the same total payoff. 

    \item The capacity-constrained efficient allocation can be implemented via linear prices equal to inflated marginal costs. 
    \end{enumerate}
\end{corollary}
Thus,  capacity constraints do not change the qualitative properties of the solution, but  lead to an inefficiency in the relative allocation of token types, and in the split between those buyer types who choose to fine-tune $(z>0)$ and those who do not $(z=0)$.

The special case where all  capacity constraints bind is of particular interest, because it is instrumental in the characterization of the buyer's demand for token budgets in \autoref{sec:packages}. When all constraints bind, the total production cost is pinned down. 
 The planner's and the buyer's solutions then coincide, i.e., the constrained efficient allocation solves the problem of a buyer who has access to inference and fine-tuning token budgets $(X,Z)$. 
 
The optimal allocation of a fixed budget $(X,Z)$ solves the following problem:

\begin{equation}\label{eq:token-budget-buyer-problem-text}
    \max_{x_{ij}\geq0} \int_0^1 w_i \Psi(x_i)\Phi(Z)\,di,\quad\mathrm{s.t.}\ \int_0^1 x_{ij}\,di=X_j \textrm{ for } j\in[J].
\end{equation}
Applying the homogeneity argument again,  $\Phi(Z)$ factors out, and the buyer optimally sets $x_i=r_i\,d$ with  $r_i\propto w_i^{1/(1-\sigma)}$.  Since the budget constraint for each token class $j$ requires $d_j\int_0^1 r_i\,di=X_j$, the common direction is $d=X/\int_0^1 r_i\,di$, which yields a simple expression for the buyer's optimal payoff.

The following result is the key step in our analysis. It shows that the buyer's indirect utility from any token bundle is multiplicatively separable in the aggregate type and an aggregate quality index, placing the seller's problem squarely in the Mussa-Rosen framework.

\begin{proposition}
[Buyer Indirect Utility]\label{lem:buyer-optimal-package}
    For any inference and fine-tuning token budgets $X=(X_1,\dots,X_J)\geq0$ and $Z=(Z_1,\dots,Z_K)\geq0$, the indirect utility of buyer type $\w$ is $$U(\w,X,Z)=\theta(\w)\, \Psi(X)\Phi(Z),$$ where $\theta(\w)$ is the aggregate type defined in \eqref{eq:CES_type}.
\end{proposition}
We therefore obtain a tractable expression for the  buyer's payoff by means of a  representative task, whose value is given by the aggregate type $\theta$, to which the buyer assigns the  entire token budget. Heterogeneity across tasks washes out, and only the aggregate matters. A notable implication is that the buyer's unobserved allocation of tokens across tasks generates no additional incentive constraints beyond those of the standard screening problem. Under homogeneity, the cost-minimizing input mix is independent of scale, so the buyer's within-bundle allocation is pinned down regardless of type. This product form is the key input to the seller's problem, which we turn to next.

\section{Menus of Token Budgets}\label{sec:packages}
The main result of this section is that the seller's optimal mechanism takes the form of a menu of committed-spend contracts: each buyer pays an upfront fee for a spending budget that she allocates freely across token classes priced at the provider's marginal cost. Higher types purchase larger budgets at higher fees, with quantity discounts that are consistent with observed industry pricing. The formal analysis proceeds in two steps: we first characterize the optimal direct mechanism (\autoref{sec:optimal-budget}) and then show that it admits intuitive indirect implementations (\autoref{sec:two-part-tariff}).

We now characterize the provider's profit-maximizing menu of token budgets.  By \autoref{lem:buyer-optimal-package}, the buyer's indirect utility from a bundle $(X,Z)$ is $\theta(\w)\,\Psi(X)\Phi(Z)$.  Two features of this expression are worth emphasizing.  First, private information enters only through the scalar $\theta$: the seller's multidimensional screening problem reduces to a one-dimensional problem. Second, the payoff is multiplicatively separable in type and allocation, placing us in the framework of \cite{mussa1978monopoly}.

\subsection{Optimal Budget Mechanism}\label{sec:optimal-budget}

Define the effective type as $\theta$ and the effective allocation as the \emph{aggregate quality} given by
\begin{equation}
    Q(X,Z)\triangleq\Psi(X)\Phi(Z).
\end{equation}
The effective production costs of producing the aggregate quality $Q$ can be found via a cost-minimization problem:
\begin{equation}\label{eq:costs_package}
    C(Q)\triangleq\min_{X,Z\geq0}\  \sum_{j=1}^J c_j X_{j} +\sum_{k=1}^K \hat{c}_k Z_k,\quad\mathrm{s.t. }\  \Psi(X)\Phi(Z)=Q.
\end{equation}
The solution to the cost-minimization  problem has the following properties.

\begin{lemma}[Cost Function]\label{lem:cost-function}
The cost function $C(Q)$ is strictly increasing and strictly convex and satisfies $C'_+(0)=0$.\footnote{In \autoref{sec:multi-model}, we show that if $\Phi(Z)$ is also a homogeneous function, then $C(Q)$ is simply a power function.} 
\end{lemma}
Convexity follows from the strict concavity of the gain function $g$: producing higher aggregate quality requires disproportionately more tokens. The property $C'_+(0)=0$ follows from the Inada conditions. Economically, $C'_+(0)=0$ means that the first unit of aggregate quality is arbitrarily cheap to produce. Combined with convexity, this implies that it is never efficient to exclude any buyer type entirely. Thus,  in the  monopoly problem,  exclusion is driven by information rents rather than production costs.

By \autoref{lem:cost-function}, the analysis of \cite{mussa1978monopoly} applies. Denote the prior distribution of $\theta$ by $F$, which is derived from $F_{\w}$. The seller chooses an aggregate quality schedule $Q(\theta)$ and a transfer schedule $t(\theta)$ to solve
\[
    \max_{Q(\cdot),\,t(\cdot)}\ \expect_F\bigl[t(\theta)-C(Q(\theta))\bigr],
\]
subject to incentive compatibility and individual rationality.

As usual, let
\begin{equation}
\MR(\theta)\triangleq \theta-\frac{1-F(\theta)}{f(\theta)}, 
\end{equation}
denote the virtual (aggregate) type. If the virtual value $\MR$ is not increasing, let  $\bar\MR$ denote its Myerson-ironed version. Then, all $\theta$ with $\bar\MR(\theta)\leq0$ are excluded. For all other $\theta$, the optimal allocation is uniquely pinned down by:
\begin{equation}\label{eq:quality_package}
    \bar\MR(\theta)=C'(Q(\theta)).
\end{equation}
Denoting the corresponding optimal transfers by 
\begin{equation}\label{eq:transfers_package}
    t(\theta)=\theta\,Q(\theta)-\int_0^\theta Q(\ind) d\ind,
\end{equation}
we can then fully characterize the optimal mechanism.

\begin{proposition}[Optimal Menu]\label{prop:packages} The optimal menu of token budgets  is $\{(X(\theta),Z(\theta),t(\theta))\}_{\theta}$, 
where  $(X(\theta),Z(\theta))$ are the cost-minimizing token budgets  that deliver quality $Q=Q(\theta)$ at prices $t(\theta)$ as defined in \eqref{eq:costs_package}-\eqref{eq:transfers_package}.
\end{proposition}

The optimal menu offers a continuum of plans.  All types $\w$ with the same aggregate type $\theta(\w)$ choose the same item. Low-type buyers are excluded. All served buyers receive distorted-downward quality (fewer tokens than the efficient allocation would prescribe) except at the very top. Buyers with higher aggregate types purchase strictly larger token budgets and pay strictly higher transfers, but enjoy quantity discounts: the average price per unit of quality falls with $\theta$.

\subsection{Indirect Mechanisms}\label{sec:two-part-tariff}

The optimal direct mechanism specifies token quantities for each type. In practice, providers do not announce menus of token quantities; they post prices and spending limits. We now show that three natural pricing formats implement the optimum.

We first consider committed-spend mechanisms, where the terms of the contract with a buyer are related to a maximum or minimum total expenditure.

\begin{definition}[Maximum-Spend Mechanism]
A maximum-spend mechanism is a collection of token prices $p$ and a menu of monetary budgets and transfers $\{(B_n,T_n)\}_n$ such that, upon selecting item $n$, the buyer pays $T_n$ for access to budget $B_n$, which he can freely spend  on tokens priced at $p$.
\end{definition}

While the optimal budget (direct) mechanism is a cap on quantities, the maximum-spend mechanism limits quality by capping expenditures at fixed prices. Those expenditures are virtual in that they do not contribute to payment beyond $T_n$, but they govern token allocation. As we will see in Section \ref{sec:application}, the maximum-spend  mechanism is used in practice, for example, by Quora's Poe (see Section~\ref{sec:application}), and it is closely related to the Cost-Based tariffs (where $p=c$) introduced in \cite{arms96}. Indeed, \autoref{prop:cost-based} in \autoref{app:alloc} derives analogous conditions to those in \cite{arms96} under which a maximum-spend mechanism is optimal across all mechanisms, including those that specify a task-by-task token allocation.

An alternative implementation does not impose a cap on spending, but exposes the buyer to variable marginal prices, much like committed-spend mechanisms in cloud computing.

\begin{definition}[Minimum-Spend Mechanism]
A minimum-spend mechanism is a menu of monetary budgets and transfers $\{(p_n,T_n)\}_n$ such that, upon selecting item $n$, the buyer commits to spending at least $T_n$ on tokens priced at  $p_n$.
\end{definition}
A minimum-spend mechanism allows buyers to commit to larger budgets (i.e., total expenditures) to unlock variable-price discounts. In contrast to a maximum-spend mechanism, payment comes from actual token consumption rather than from an upfront payment.

\begin{definition}[Two-Part-Tariff Mechanism]
A two-part-tariff mechanism is a menu of prices and transfers $\{(p_n,T_n)\}_n$ such that, upon selecting item $n$, the buyer pays $T_n$ and can buy any number of tokens priced at  $p_n$.
\end{definition}

A two-part tariff is a classical mechanism that combines upfront and consumption payments but does not impose any token consumption limits.

Define a type-dependent \emph{markup} $m(\theta)$ as:
\begin{equation}
    m(\theta)\triangleq \frac{\theta}{\MRI(\theta)}.
\end{equation}
The markup $m(\theta)$ is the ratio of the true type to the virtual type; it captures the information-rent wedge that the seller imposes. Higher markups on lower types reflect greater information rents extracted from higher types in the menu.

\begin{proposition}[Indirect Implementation]\label{prop:indirect-implementation}
An  optimal menu of token budgets can be implemented via a maximum-spend mechanism. If $m(\theta)$ is decreasing (e.g., if $F$ has an increasing hazard rate), then an  optimal menu of token budgets can be implemented via  a minimum-spend mechanism and a two-part tariff mechanism.
\end{proposition}

Implementability by a maximum-spend mechanism is straightforward
and relies on the fact that the optimal token allocation is
constrained-efficient, as in \cite{dodkt25}: it suffices to set
prices equal to marginal costs and to choose token expenditures
and transfers so as to mimic those under the direct mechanism.

Implementability by minimum-spend and two-part tariff mechanisms 
requires more care.  Because $\Psi$ is homogeneous, the 
cost-minimizing input mix is the same for every quality level 
$Q$; only the scale changes.  A two-part tariff that preserves 
this input mix must therefore set token prices proportional to 
marginal costs; any other price vector would distort the buyer's 
allocation away from cost minimization.  The proportionality 
factor is the markup $m(\theta)=\theta/\MRI(\theta)$, which 
inflates all marginal costs equally.  The condition that 
$m(\theta)$ is decreasing ensures that higher types, who select 
larger budgets, face lower per-token prices, so the menu is 
incentive-compatible.  In the minimum-spend implementation, the 
markup  equals the ratio between total revenue and 
total cost for type $\theta$. Section~\ref{sec:application} shows that each of these three implementations corresponds to an observed pricing format at leading platforms.
\section{Multiple Models and Competition} \label{sec:multi-model}

In practice, every major LLM provider offers multiple models that differ in capability and cost (e.g., Anthropic's Haiku, Sonnet, and Opus, or OpenAI's GPT-4o-mini and o1). A buyer can assign different tasks to different models, and the provider can screen on both usage quantity and model access. This section extends our framework to this richer environment. Two new questions arise: How does a buyer optimally allocate tasks across models? And how does a profit-maximizing provider design menus that screen on both dimensions?

Specifically, we extend our setting and allow there to be $L$ models, each with a model-specific gain function:
\begin{equation}\label{eq:gain-multiple}
g_l(x_{i},z)=\Psi_l(x_i)\Phi_l(z),
\end{equation} 
where $\Psi_l$ and $\Phi_l$ are assumed to have the same properties as in the baseline model, with $\Psi_l$ being homogeneous of degree $\sigma_l$. We order the models so that $\sigma_1\leq\sigma_2\leq\dots\leq\sigma_L$. The token costs are model-specific $(c_l,\hat c_l)$.

 The buyer can process different tasks with different models; however, he cannot use two models for the same task.\footnote{Equivalently, we could assume that the buyer can combine tokens from two models on any task, with a performance given by  the sum of individual gain functions.} If the buyer of type $\w$ processes tasks $i\in I_l$ with model $l$, his total payoff ignoring the payment is: 
\begin{equation}
   \sum_{l=1}^L \int_{I_l} w_i  g_l(x_{li},z_l) di.
\end{equation}

In this setting, a bundle specifies token budgets  for all available models  $(X_l,Z_l)_{l=1}^L=(X_{l1},\dots,X_{lJ},Z_{l1},\dots,Z_{lK})_{l=1}^L$.

As a preliminary step for the subsequent analysis, we show that the buyer's value for a bundle $(X_l,Z_l)_{l=1}^L$ depends only on the aggregate quality of each model, 
\begin{equation}
Q_l\triangleq g_l(X_l,Z_l),
\end{equation}
and admits a tractable closed-form expression. To this end, we fix $\w$ and  order the tasks so that $w_i$ is increasing. We assume that $Q_l>0$ for all $l$ (if $Q_l=0$ then the $l$-th model can be ignored). If the buyer allocates tasks in the set $I_l$ to model $l$, then his payoff from model $l$ is, by the arguments of \autoref{lem:buyer-optimal-package},
\begin{equation}\label{eq:buyer-optimal-package-l}
U_l(I_l,Q_l)=\theta_l(I_l) Q_l,     
\end{equation}
where $\theta_l(I_l)\triangleq (\int_{I_l} w_i^{1/(1-\sigma_l)}di)^{1-\sigma_l}$. The total payoff from the task-model allocation $(I_l)_{l=1}^L$ is
\begin{equation}
    \sum_{l=1}^L U_l(I_l,Q_l)=\sum_{l=1}^L \theta_l(I_l) Q_l.
\end{equation}
The optimal payoff $U^*(Q_1,\dots,Q_L)=\sum_{l=1}^L U_l(I^*_l,Q_l)$ is evaluated at the optimal task-model allocation $(I^*_l)_{l=1}^L$.

\begin{lemma}[Buyer-Optimal Payoff]\label{lem:buyer-optimal-multiple} If $\sigma_l=\sigma$ for all $l\in[L]$, then
\begin{equation}\label{eq:buyer-value-multimodel}
    U^*(Q_1,\dots,Q_L)=\theta\left(\sum_{l=1}^L Q_l^{1/\sigma}\right)^\sigma.
\end{equation}
\end{lemma}

By \autoref{lem:buyer-optimal-multiple}, despite the combinatorial complexity of assigning a continuum of tasks to multiple models, the buyer's payoff from a bundle of tokens across different models admits a simple product decomposition into an aggregate type and a CES-style aggregate quality over models. The elasticity of substitution across models is pinned down by the common returns-to-scale parameter $\sigma$, and the single-model tractability of Sections~3--4 extends to the multi-model environment.

\subsection{Efficient Solution}

We first compute the efficient allocation in this setting. Since the buyer's value takes a simple form (\ref{eq:buyer-value-multimodel}) as a function of aggregate qualities of each model, the efficient allocation delivers the efficient amount of these qualities in a cost-efficient way.

\begin{assumption}[Homogeneous Fine-Tuning]\label{ass:homogeneous-fine-tuning} For each $l=1,...,L$:
\begin{enumerate}
    \item  $\Psi_l$ is homogeneous of degree $\sigma\in(0,1)$. 
    \item $\Phi_l$ is homogeneous of degree $\hat\sigma_l<1-\sigma$. 
\end{enumerate}
     
\end{assumption}
All subsequent analysis holds under \autoref{ass:homogeneous-fine-tuning}, so we omit it from the statements. \autoref{ass:homogeneous-fine-tuning} allows us to obtain closed-form expressions because of the following result:

\begin{lemma}[Model-Specific Cost Function]\label{lem:homog-indirect-cost}
The  cost function of delivering aggregate quality $Q_l$ through model $l$ is $C_l(Q_l)=\kappa_l Q_l^{1/(\sigma+\hat{\sigma}_l)}$ for some constant $\kappa_l>0$.
\end{lemma}

By \autoref{lem:homog-indirect-cost}, the minimal cost of obtaining quality $Q$ from $L$ models is
\begin{equation*}
    C(Q)=\min_{Q_1,\dots,Q_L\geq0} \sum_{l=1}^L \kappa_l Q_l^{1/(\sigma+\hat{\sigma}_l)},\quad\mathrm{s.t.}\ \sum_{l=1}^L Q_l^{1/\sigma}=Q^{1/\sigma}.
\end{equation*}
The optimal solution utilizes a single model, resulting in
\begin{equation}\label{eq:cost-multimodel}
    C(Q)=\min_{l\in[L]}  \kappa_l Q^{1/(\sigma+\sigmah_l)} .
\end{equation}
Thus, the rich functional model heterogeneity (\ref{eq:gain-multiple}) reduces to heterogeneity in two parameters $(\kappa_l,\hat{\sigma}_l)$ that roughly correspond to  ``cost-effectiveness''  and ``propensity to fine-tune.'' Heterogeneity in $\kappa_l$ is vertical: higher $\kappa_l$ correspond to overall costlier models. Heterogeneity in $\sigmah_l$ is horizontal: models with high $\hat{\sigma}_l$ are better at generating high $Q$ whereas models with low $\hat{\sigma}_l$ are better at generating low $Q$. As a result, higher values of $Q$ are optimally produced by models with higher $\hat{\sigma}_l$. In other words, models with higher returns to fine-tuning have flatter cost curves and are cheaper for producing high qualities, introducing a natural form of horizontal differentiation among models.

Notably, the cost function $C(Q)$ is not convex but piecewise convex, with kinks at the model switches, which translates into unusual properties of the efficient and profit-maximizing allocations. Specifically, the efficient quality allocation for type $\theta$ when using model $l$ solves:
\begin{equation*}
     \max_{Q \geq0}\ \theta Q- \kappa_l Q^{1/(\sigma+\hat{\sigma}_l)},
\end{equation*}
leading to the following characterization:
\begin{proposition}[Efficient Allocation]\label{prop:efficient-multiple}
In the efficient allocation, each type $\theta$ processes all tasks with a single model:
\begin{equation}\label{eq:l-eff-multimodel}
    l^{*}\in\arg\max_l\ (1-\sigma-\sigmah_l)\left(\frac{  (\sigma+\sigmah_l)}{\kappa_l}\right)^{(\sigma+\sigmah_l)/(1-\sigma-\sigmah_l)}\theta^{1/(1-\sigma-\sigmah_l)},
\end{equation}
with the aggregate quality: 
\begin{equation}\label{eq:q-eff-multimodel}
    Q^{*}(\theta)=\left(\frac{ \theta(\sigma+\sigmah_{l^{*}})}{\kappa_{l^{*}}}\right)^{(\sigma+\sigmah_{l^{*}})/(1-\sigma-\sigmah_{l^{*}})}.
\end{equation}
The efficient token allocation is given by \autoref{lem:homog-indirect-cost} evaluated at $Q^{*}$ from (\ref{eq:q-eff-multimodel}).
\end{proposition}

Every type uses a single model. Higher aggregate types $\theta$ consume higher aggregate qualities by using models with higher fine-tuning capabilities $\hat{\sigma}_l$, i.e., $Q^{*}(\theta)$ is increasing. 

Because $C(Q)$ is not continuously differentiable, the allocation is discontinuous at the model switch. Indeed, it follows from (\ref{eq:q-eff-multimodel}) and (\ref{eq:l-eff-multimodel}) that if the efficient allocation at type $\theta$ is indifferent between $Q_l$ of model $l$ and $Q_{l'}$ of model $l'$ such that $\sigmah_{l'}>\sigmah_l$, then $$\frac{Q_{l'}}{Q_l}=\frac{1-\sigma-\sigmah_l}{1-\sigma-\sigmah_{l'}}>1.$$

\subsection{Multi-Model Monopolist}
We now assume that all $L$ models are sold by a monopolist who offers a menu. Each item specifies a transfer and a collection of token budgets for different models  $(X_l,Z_l)_{l=1}^L$.

By \autoref{lem:buyer-optimal-multiple}, the seller's problem is equivalent to the optimal screening problem in \cite{mussa1978monopoly}, in which the buyer's type is $\theta\in\reals$, the seller chooses a menu of $(Q,t)$, the buyer's payoff is $U(\theta,Q)=\theta Q$, and the cost of obtaining a quality $Q$ is equal to $C(Q)$ as in (\ref{eq:cost-multimodel}). Although $C(Q)$ is not convex, the Envelope theorem applies and this problem is equivalent to a maximization of virtual surplus:
\begin{equation}
    \max_{Q(\cdot):\mathrm{\,increasing}} \int_0^1 \Bigl(\MR(\theta)Q(\theta)-C(Q(\theta))\Bigr)dF(\theta),
\end{equation}
where $\MR(\theta)=\theta-(1-F(\theta))/f(\theta)$. 

Assuming $\MR(\theta)$ is increasing, the solution can be obtained pointwise to equal an efficient allocation with respect to the virtual type: $Q^{\mathrm{m}}(\theta)=Q^{*}(\MR(\theta))$. By \autoref{prop:efficient-multiple}, if $\MR(\theta)<0$, then the buyer is excluded, $Q^{\mathrm{m}}=0$; if $\MR(\theta)\geq0$,
\begin{align}\label{eq:multimodel-opt-l}
    &l^{\mathrm{m}}(\theta)\in\arg\max_l\ (1-\sigma-\sigmah_l)\left(\frac{  (\sigma+\sigmah_l)}{\kappa_l}\right)^{(\sigma+\sigmah_l)/(1-\sigma-\sigmah_l)}\MR(\theta)^{1/(1-\sigma-\sigmah_l)},\\
    &Q^{\mathrm{m}}(\theta)=\left(\frac{ \MR(\theta)(\sigma+\sigmah_{l^{\mathrm{m}}})}{\kappa_{l^{\mathrm{m}}}}\right)^{(\sigma+\sigmah_{l^{\mathrm{m}}})/(1-\sigma-\sigmah_{l^{\mathrm{m}}})}.\label{eq:multimodel-opt-Q}
\end{align}
Since $Q^{*}$ is increasing, the pointwise solution solves the problem. Optimal transfers are 
\begin{equation}\label{eq:multimodel-opt-T}
  t^{\mathrm{m}}(\theta)=\theta Q^{\mathrm{m}}(\theta)-\int_0^\theta Q^{\mathrm{m}}(\ind)d\ind .   
\end{equation}
As under the efficient allocation, because $C(Q)$ is not continuously differentiable,  the quality and transfer schedules are discontinuous.
\begin{proposition}[Multi-Model Monopoly]\label{prop:packages-multiple} If $\MR(\theta)$ is increasing, then the optimal token-budget menu is given by $\{((X_l(\theta),Z_l(\theta))_{l=1}^L,t(\theta))\}_{\theta}$,
where  $(X_l(\theta),Z_l(\theta))_{l=1}^L$ are efficient token budgets that deliver quality $Q^{\mathrm{m}}(\theta)$ at prices  $t(\theta)$ as defined in (\ref{eq:multimodel-opt-Q})  and (\ref{eq:multimodel-opt-T}).  
\end{proposition}

The monopolist assigns each buyer type to a single model, with higher types using more capable models. All types $\w$ with the same aggregate type $\theta(\w)$ choose the same item and use one model, $l^{\mathrm{m}}(\theta)$ as defined in (\ref{eq:multimodel-opt-l}), for all tasks. The quality schedule exhibits discrete jumps at model-switching thresholds: a buyer at the margin between two models jumps to a strictly higher quality when upgrading. OpenAI's tier structure, which reserves its most compute-intensive reasoning model for the highest-paying subscribers, provides a direct empirical counterpart; see the discussion in Section~\ref{sec:application}.

\subsection{Leader-Fringe Competition}
We now study  competition between a proprietary leader and an open-source competitive fringe. The key economic forces are: (i) the fringe provides an outside option whose value depends on the buyer's type; (ii) for intermediate types, the leader must provide enough tokens to deter fringe top-up; (iii) for high types, the leader acts as an unconstrained monopolist. The resulting allocation has three distinct regions, generating a richer pattern of distortions than either the single-model monopoly or the multi-model monopoly.

Specifically, we assume that there is a single leader, corresponding to the highest-capability model ($l=L$) in the previous section, and a continuum of firms in a competitive fringe that sell their tokens at fixed per-token prices equal to marginal costs. We assume that the leader possesses a proprietary model characterized by the aggregate cost parameter $c_L>0$, returns to intensity $\sigma_L=\sigma>0$, and returns to fine-tuning $\hat{\sigma}_L>0$, such that $\sigma+\hat{\sigma}_L<1$. The fringe possesses an open-source model characterized by the lower aggregate cost parameter $c_F\in[0,c_L)$, the same returns to intensity  $\sigma_F=\sigma$, and lower returns to fine-tuning $\hat{\sigma}_F\in[0,\hat{\sigma}_L)$.

The buyer has a private type $\w$. By \autoref{lem:buyer-optimal-multiple}, only the aggregate type $\theta(\w)$ matters, so we take $\theta$ as a primitive and assume that $\theta$ is distributed according to $F$ with strictly positive density $f$ everywhere on $[0,1]$. The buyer can multi-home and can buy at most one item from the leader. 

We want to solve the leader's problem, which can be viewed as a monopolistic design of an optimal menu $(q,t)$  subject to the competitive pressure from the fringe. For notational simplicity, define the \emph{quantities}:
\begin{align*}
&q_L\triangleq Q_L^{1/(\sigma+\hat{\sigma}_L)}=g_L(X_{L1},\dots,X_{LJ},Z_{L1},\dots,Z_{LK})^{1/(\sigma+\hat{\sigma}_L)},\\
&q_F\triangleq Q_F^{1/(\sigma+\hat{\sigma}_F)}=g_F(X_{F1},\dots,X_{FJ},Z_{F1},\dots,Z_{FK})^{1/(\sigma+\hat{\sigma}_F)}.
\end{align*}

The payoff of type $\theta$ from having purchased quantity $q_L$ from the leader is
\begin{equation}
\max_{q_F\geq 0}\ \theta (q_L^{(\sigma+\hat{\sigma}_L)/\sigma}+q_F^{(\sigma+\hat{\sigma}_F)/\sigma})^\sigma-c_F q_F,
\end{equation}
with a (type-dependent) outside option corresponding to $q_L=0$. The first-order condition for the  buyer's  problem of purchasing fringe quantity is
\begin{equation}\label{eq:fringe-foc-1}
    \theta (\sigma+\hat{\sigma}_F)(q_L^{(\sigma+\hat{\sigma}_L)/\sigma}+q_F^{(\sigma+\hat{\sigma}_F)/\sigma})^{\sigma-1}q_F^{\hat{\sigma}_F/\sigma}=c_F.
\end{equation}

In general, (\ref{eq:fringe-foc-1}) does not admit a closed-form solution. It does so in the case of $\hat{\sigma}_F=0$. Specifically, define
\begin{equation}\label{eq:qhat}
    \qh(\theta)\triangleq \left(\frac{\theta \sigma}{c_F}\right)^{\sigma/((1-\sigma)(\sigma+\hat{\sigma}_L))},\quad \psi(\theta)\triangleq \theta^{\frac{1}{1-\sigma}}(1-\sigma)\left(\frac{\sigma}{c_F}\right)^{\sigma/(1-\sigma)}.
\end{equation}
If the  quantity purchased from the leader is $q_L<\qh(\theta)$, the  buyer will purchase fringe tokens to achieve the optimal total quantity $\qh(\theta)$; if $q_L>\qh(\theta)$, the  buyer will single-home with the leader. The buyer's outside option utility corresponding to $q=0$ is  $\psi(\theta)$. We then obtain the following characterization of the buyer's demand for the leader's quantity.
\begin{lemma}[Buyer's Payoff]\label{lemma:buyer_pay}
The leader's problem is equivalent to one in which the buyer has a zero outside option and the following payoff:
\begin{align}
    u(\theta,q)&=\begin{cases}
c_F q^{(\sigma+\hat{\sigma}_L)/\sigma}, & \textrm{if\ }q<\qh(\theta),\\
\theta q^{\sigma+\hat{\sigma}_L}-\psi(\theta),& \textrm{if\ }q\geq\qh(\theta).
\end{cases}\label{eq:payoff_norm}
\end{align}
\end{lemma}
The payoff $u(\theta,q)$ in \autoref{lemma:buyer_pay} is continuous, differentiable, and increasing in both parameters. In the region $q<\qh(\theta)$, it is convex in $q$ and independent of $\theta$. In the region $q>\qh(\theta)$, it is concave in $q$ and supermodular in $q$ and $\theta$. 

The two-regime structure has a clear economic interpretation. In the first regime ($q<\qh(\theta)$), the buyer supplements leader tokens with fringe tokens to achieve the optimal total quantity, so her marginal value of additional leader tokens is pinned down by the fringe price $c_F$; the leader faces a perfectly elastic residual demand. In the second regime ($q\geq\qh(\theta)$), the buyer single-homes with the leader, and her value is the standard concave payoff minus the outside option $\psi(\theta)$ she forgoes by not using the fringe. The transition between regimes is where the leader's competitive constraint binds.

The leader's problem is an instance of screening with a type-dependent outside option \citep{jull00}; the reformulation (\ref{eq:payoff_norm}) absorbs this outside option into the buyer's payoff. An optimal menu can then be characterized by standard methods.\footnote{See \cite{cade13} for a related analysis of nonlinear competition between a dominant firm and a competitive fringe under private information about demand.} Indeed, $u(\theta,q)$ satisfies the  (weak) single-crossing constraints: $u_{\theta q}(\theta,q)=0$ if $q<\qh(\theta)$ and $u_{\theta q}(\theta,q)>0$ if $q\geq\qh(\theta)$. Allocation $q(\theta)$ is implementable   only if it is increasing whenever $q\geq\qh(\theta)$, and the profit associated with allocation $q$ is: 
\begin{equation}\label{eq:virtual-surplus-fringe}
    \Pi(q)=\int_0^1 \left(u(\theta,q(\theta))-c_L q(\theta)-\frac{1-F(\theta)}{f(\theta)} u_\theta(\theta,q(\theta))\right)f(\theta)d\theta.
\end{equation}

In regular environments, this profit can be maximized pointwise. To this end, denote by $q^{\mathrm{m}}(\theta)$ the optimal monopoly quantity in the absence of the competitive fringe
\begin{equation}\label{eq:qint}
    q^{\mathrm{m}}(\theta)=\left(\frac{(\sigma+\hat{\sigma}_L)\MR(\theta)}{c_L}\right)^{1/(1-\sigma-\hat{\sigma}_L)}.
\end{equation}

\begin{proposition}[Leader-Fringe Competition]\label{prop:fringe}
Let $F$ satisfy a monotone hazard rate and $\hat{\sigma}_F=0$. Then, there exist $\underline\theta$ and $\overline\theta$, $\underline\theta\leq \overline\theta$, such that (i) for $\theta\leq\underline\theta$,  $q(\theta)=0$, (ii) for $\theta\in(\underline\theta,\overline\theta)$,  $q(\theta)=\qh(\theta)$ given by (\ref{eq:qhat}), and (iii) for $\theta\geq \overline\theta$,  $q(\theta)=q^{\mathrm{m}}(\theta)$ given by (\ref{eq:qint}).
\end{proposition}

Under the optimal mechanism, low types  purchase exclusively from the fringe. All higher types purchase exclusively from the leader. Out of those, the midline types are on the margin of whether to buy from the fringe, whereas the highline types strictly prefer to purchase from the leader. Depending on parameters, the midline region may not exist, $\underline\theta=\overline\theta$, or the highline region may not exist, $\overline\theta>1$.

The three-region structure reflects three distinct economic regimes. In the \emph{fringe-only} region ($\theta\leq\underline\theta$), the buyer's willingness to pay for the leader's superior fine-tuning capability is too low to justify adoption. In the \emph{deterrence} region ($\underline\theta<\theta<\overline\theta$), the leader offers exactly enough tokens to make the buyer indifferent between single-homing with the leader and supplementing with the fringe; this pins the leader's quantity to $\qh(\theta)$, which rises with type. In the \emph{monopoly} region ($\theta\geq\overline\theta$), the buyer's value is high enough that fringe competition no longer constrains the leader, who reverts to the standard downward-distorted monopoly quantity $q^{\mathrm{m}}(\theta)$. The novel region is the deterrence region, which has no analog in either the single-model monopoly or the multi-model monopoly.

\paragraph{Comparison with Efficient Allocation}
It is instructive to compare the allocation under leader-fringe competition with the efficient benchmark. By \autoref{prop:efficient-multiple}, defining
\begin{equation*}
    u_l^*(\theta)=(1-\sigma_l-\hat{\sigma}_l)\theta^{1/(1-\sigma_l-\hat{\sigma}_l)}\left(\frac{\sigma_l+\hat{\sigma}_l}{c_l}\right)^{(\sigma_l+\hat{\sigma}_l)/(1-\sigma_l-\hat{\sigma}_l)},
\end{equation*}
type $\theta$ uses the fringe model if and only if $u^*_F(\theta)>u^*_L(\theta)$. The indifference type at which $u^*_F(\theta)=u^*_L(\theta)$ is (recall that $\sigma_L=\sigma_F=\sigma$ and $\hat{\sigma}_F=0$)
\begin{equation}\label{eq:eff_threshold}
    \hat\theta=\left(\frac{1-\sigma}{1-\sigma-\hat{\sigma}_L}\right)^{(1-\sigma-\hat{\sigma}_L)(1-\sigma)/\hat{\sigma}_L}\left(\frac{\sigma}{c_F}\right)^{\sigma(1-\sigma-\hat{\sigma}_L)/\hat{\sigma}_L}\left(\frac{c_L}{\sigma+\hat{\sigma}_L}\right)^{(\sigma+\hat{\sigma}_L)(1-\sigma)/\hat{\sigma}_L}.
\end{equation}
If $\theta<\hat\theta$, then the type uses the fringe model: $q_F(\theta)=(\theta\sigma/c_F)^{1/(1-\sigma)}$ and $q_L(\theta)=0$.  If $\theta\geq\hat\theta$, then the type uses the leader model: $q_F(\theta)=0$ and $q_L(\theta)=(\theta(\sigma+\hat{\sigma}_L)/c_L)^{1/(1-\sigma-\hat{\sigma}_L)}$.

Comparing this with the fringe-competition allocation characterized in \autoref{prop:fringe}, we see two main differences: First, we see that the ``highline'' leader quantity  $q^{\mathrm{m}}(\theta)$ is distorted downward from efficiency  because $\MR(\theta)$ is smaller than $\theta$. This is natural, since in that range the leader is unconstrained by the fringe and acts as a monopolist. Second, there is an additional \emph{extensive margin distortion}, because $\hat\theta<\underline{\theta}$.\footnote{Indeed,
$
    \frac{\hat\theta}{\underline{\theta}}=\big[\big( \frac{1-\sigma}{1-\sigma-\hat{\sigma}_L}\big)^{1-\sigma-\hat{\sigma}_L}\big( \frac{\sigma}{\sigma+\hat{\sigma}_L}\big)^{\sigma+\hat{\sigma}_L}\big]^{(1-\sigma)/\hat{\sigma}_L},
$
and thus, by the strict concavity of the logarithm and Jensen's inequality,
$
    \ln\left(\frac{\hat\theta}{\underline{\theta}}\right)=\frac{1-\sigma}{\hat{\sigma}_L}\big[(1-\sigma-\hat{\sigma}_L)\ln\big(\frac{1-\sigma}{1-\sigma-\hat{\sigma}_L}\big)+(\sigma+\hat{\sigma}_L)\ln\big(\frac{\sigma}{\sigma+\hat{\sigma}_L}\big)\big]< 0.
$}

\paragraph{Comparison with Multi-Model Monopoly} Another natural benchmark is the case of a monopolist possessing the leader and fringe models. By \autoref{prop:packages-multiple}, the monopolist effectively faces a Mussa-Rosen problem with the buyer's payoff:
\begin{equation*}
    u(\theta,q_L,q_F)=\theta (q_F+q_L^{(\sigma+\hat{\sigma}_L)/\sigma})^\sigma.
\end{equation*}
and  will be supplying an efficient allocation with respect to the virtual type. If $\theta<\theta^{\mathrm{m}}$, where $\MR(\theta^{\mathrm{m}})\triangleq\hat\theta$ and $\hat\theta$ is defined at (\ref{eq:eff_threshold}), then the type uses the fringe model and $q_F(\theta)=(\MR(\theta)\sigma/c_F)^{1/(1-\sigma)}$ and $q_L(\theta)=0$.  If $\theta\geq\theta^{\mathrm{m}}$, then the type uses the leader model with $q_F(\theta)=0$ and $q_L(\theta)=(\MR(\theta)(\sigma+\hat{\sigma}_L)/c_L)^{1/(1-\sigma-\hat{\sigma}_L)}$. 

Relative to the efficiency benchmark, the allocation is distorted downward within each model, except at the very top. In addition, fewer types consume the leader model.
 
Relative to the leader-fringe environment of \autoref{prop:fringe}, the integrated multi-model monopolist internalizes the fringe technology and therefore does not face a competitive outside option. This has two implications. First, the fringe allocation is no longer pinned down by marginal-cost pricing: types that use the fringe model receive
$q_F(\theta)=(\MR(\theta)\sigma/c_F)^{1/(1-\sigma)}$ (and are excluded whenever $\MR(\theta)<0$),
whereas under leader-fringe competition the corresponding types purchase the efficient quantity
$(\theta\sigma/c_F)^{1/(1-\sigma)}$ from the fringe. Second, the ``midline'' region
$\theta\in(\underline{\theta},\overline{\theta})$ in which the leader supplies $q=\qh(\theta)$ to deter top-up purchases from the fringe disappears. Because the monopolist controls both models, she uses the fringe model directly as the low-quality product in the Mussa-Rosen menu and assigns each type to a single model with a single cutoff $\theta^{\mathrm m}$. Finally,  once the fringe constraint is slack, the intensive-margin provision coincides and equals the monopoly quantity $q^{\mathrm m}(\theta)$; differences between the two environments are therefore concentrated among low and intermediate types.\footnote{\cite{cade15} show that a dominant firm with a competitive advantage can profitably impose exclusive dealing on privately informed buyers; the mechanism here is analogous, though exclusivity arises from the optimal nonlinear tariff rather than from contractual restrictions.}

\autoref{fig:example-uniform-1} illustrates the optimal allocation and distortions in a uniform example. All calculations for this example are in \autoref{app:proofs}. 

The top panel plots the leader quantity $q(\theta)$, and  the bottom panel plots the fringe quantity $q_F(\theta)$. Four cutoffs, ordered $\hat\theta < \underline\theta < \theta^{\mathrm{m}} < \overline\theta$, partition the type space. The efficient allocation (solid) features a single switch at $\hat\theta$: types below $\hat\theta$ use only the fringe, types above use only the leader. The leader--fringe allocation (dotted) shifts this switch rightward to $\underline\theta$, reflecting the extensive-margin distortion. Between $\underline\theta$ and $\overline\theta$, the leader supplies the deterrence quantity $\hat{q}(\theta)$, which lies strictly below the efficient leader quantity. Above $\overline\theta$, the fringe constraint is slack and the leader reverts to the monopoly quantity $q^{\mathrm{m}}(\theta)$, which coincides with the integrated monopolist's allocation (dashed). In the bottom panel, the efficient and leader--fringe curves for $q_F$ coincide over their respective ranges, since the fringe prices at marginal cost in both cases; the difference is that fringe usage persists up to $\underline\theta$ rather than $\hat\theta$. The integrated monopoly distorts fringe provision downward and switches to the leader at the later cutoff $\theta^{\mathrm{m}}$.
\def\thhat{0.215166}
\def\thone{0.279508}
\def\thM{0.607583}
\def\thtwo{0.621614}
\begin{figure}[t!]
\begin{tikzpicture}
\begin{groupplot}[
  group style={
    group size=1 by 2,
    vertical sep=10mm,
    xlabels at=edge bottom,
    xticklabels at=edge bottom,
  },
  width=13cm,
  xmin=0, xmax=1,
  domain=0:1, samples=300,
  axis lines=left,
  tick label style={/pgf/number format/fixed},
  legend style={draw=none,fill=none,font=\small},
]
\nextgroupplot[
  height=7cm,
  ymode=log,
  ymin=1, ymax=1500,
   xlabel={$\theta$},
  ylabel={$q(\theta)$},
   xlabel style={
  at={(axis description cs:1,0)},
  anchor=north west
},
ylabel style={
rotate=-90,
  at={(axis description cs:0,0.95)},
  anchor=south east
},
  legend pos=south east,
]
\addplot[very thick,restrict y to domain=0:50,domain=\thhat:1] {(6*x)^4};
\addlegendentry{Efficient}
\addplot[very thick,forget plot] coordinates {(0,0) (\thhat,0)};
\addplot[thick,dashed,restrict y to domain=0:50,domain=\thM:1] {(6*(2*x-1))^4};
\addlegendentry{Monopoly}
\addplot[thick,dashed,forget plot] coordinates {(0,0) (\thM,0)};
\addplot[thick,dotted,domain=\thone:\thtwo] {pow(5*x,4/3)};
\addplot[thick,dotted,restrict y to domain=0:50,domain=\thtwo:1] {(6*(2*x-1))^4};
\addlegendentry{Leader--fringe}
\addplot[thick,dotted,forget plot] coordinates {(0,0) (\thone,0)};
\draw[densely dashed] (rel axis cs:\thhat,0) -- (rel axis cs:\thhat,0.9);
\draw[densely dashed] (rel axis cs:\thone,0) -- (rel axis cs:\thone,0.9);
\draw[densely dashed] (rel axis cs:\thM,0)   -- (rel axis cs:\thM,0.9);
\draw[densely dashed] (rel axis cs:\thtwo,0) -- (rel axis cs:\thtwo,0.9);
\node[anchor=south,inner sep=1pt]      at (axis cs:\thhat,840) {$\hat\theta$};
\node[anchor=south west,inner sep=1pt] at (axis cs:\thone,800) {$\underline{\theta}$};
\node[anchor=south east,inner sep=1pt] at (axis cs:\thM,800)   {$\theta^{\mathrm{m}}$};
\node[anchor=south west,inner sep=1pt] at (axis cs:\thtwo,800) {$\overline{\theta}$};
\nextgroupplot[
  height=7cm,
  ymode=log,
  ymin=1e-2, ymax=9.5,
  xlabel={$\theta$},
  ylabel={
    $q_F(\theta)$
  },
   xlabel style={
  at={(axis description cs:1,0)},
  anchor=north west
},
ylabel style={
rotate=-90,
  at={(axis description cs:0,0.95)},
  anchor=south east
}
]
\def\thhat{0.215166}   
\def\thone{0.279508}   
\def\thM{0.607583}     
\def\thtwo{0.621614}   
\addplot[very thick, domain=0:\thhat] {pow(5*x,2)};
\addlegendentry{Efficient}
\addplot[very thick, forget plot] coordinates {(\thhat,0) (1,0)};
\addplot[thick, dashed, domain=0.5:\thM] {pow(5*(2*x-1),2)};
\addlegendentry{Monopoly}
\addplot[thick, dashed, forget plot] coordinates {(0,0) (0.5,0)};
\addplot[thick, dashed, forget plot] coordinates {(\thM,0) (1,0)};
\addplot[thick, dotted, domain=0:\thone] {pow(5*x,2)};
\addlegendentry{Leader--fringe}
\addplot[thick, dotted, forget plot] coordinates {(\thone,0) (1,0)};
\draw[densely dashed] (rel axis cs:\thhat,0) -- (rel axis cs:\thhat,0.9);
\draw[densely dashed] (rel axis cs:\thone,0) -- (rel axis cs:\thone,0.9);
\draw[densely dashed] (rel axis cs:\thM,0)   -- (rel axis cs:\thM,0.9);
\node[anchor=south,inner sep=1pt] at (rel axis cs:\thhat,0.92) {$\hat\theta$};
\node[anchor=south west,inner sep=1pt] at (rel axis cs:\thone,0.91) {$\underline{\theta}$};
\node[anchor=south east,inner sep=1pt] at (rel axis cs:\thM, 0.91) {$\theta^{\mathrm{m}}$};
\end{groupplot}
\end{tikzpicture}
    \caption{Leader-Fringe allocations across different regimes. Example with $\theta\sim U[0,1]$, $\sigma=1/2$, $\hat{\sigma}_L=1/4$, $c_F=1/10$, $c_L=1/8$.}
    \label{fig:example-uniform-1}
\end{figure}

\section{LLM Pricing in Practice}\label{sec:application}

A striking feature of current LLM pricing is that different market segments implement  each of the  optimal  mechanisms   in our theoretical framework: 
consumer subscriptions implement the nonlinear menus of Sections~\ref{sec:packages} and~\ref{sec:multi-model}; model aggregators implement the committed-spend mechanisms of \autoref{sec:two-part-tariff}; and developer APIs implement the constrained-efficient linear pricing of \autoref{cor:capacity-efficient}. 

Our main findings are as follows. 
The two largest proprietary providers, Anthropic and OpenAI, offer remarkably similar price points (\$20 and \$200) but implement fundamentally different screening mechanisms. Anthropic screens through quantity alone, holding model access constant across paid tiers, matching the token-budget mechanism of \autoref{prop:packages}. OpenAI screens through both quantity and model access, reserving its most compute-intensive reasoning model for the highest tier, matching the multi-model menu of \autoref{prop:packages-multiple}. This contrast illustrates the distinction between the single-model screening of Section~\ref{sec:packages} and the multi-model versioning of Section~\ref{sec:multi-model}. Model aggregators, which resell access to upstream providers through a 
single interface, implement the committed-spend mechanisms of 
\autoref{sec:two-part-tariff}: Quora's Poe enforces a hard budget 
cap (maximum spend), while GitHub Copilot allows overage 
purchasing at linear prices (minimum spend). Finally, API pricing 
across all major providers is linear in tokens with no volume 
discounts, resembling the constrained-efficient benchmark of 
\autoref{cor:capacity-efficient} rather than any profit-maximizing 
mechanism, consistent with providers prioritizing market share 
over rent extraction in the developer segment.

\subsection{Anthropic}

Anthropic's consumer pricing for Claude best illustrates the token-budget mechanism of Section~4. Anthropic offers the same model family across all paid tiers, with differentiation occurring through usage allocations. As in the  optimal menu characterized in \autoref{prop:packages}, users  with higher aggregate types~$\theta$ select items with larger token budgets while consuming the same underlying technology.

\autoref{fig:claude-pricing} summarizes the  structure of  paid tiers as of January 2026. All paid tiers, namely Pro (\$20/month), Max~5x (\$100/month), and Max~20x (\$200/month), grant access to Haiku~4.5, Sonnet~4.5, and Opus~4.5. The key distinction is the quantity of compute: Anthropic measures usage limits in computational intensity. For example, Opus queries consume resources approximately $5\times$ faster than Sonnet. This aligns with \autoref{prop:packages}, where the optimal mechanism defines users' budgets in total tokens consumed across classes. Although Anthropic offers multiple models, all paid tiers grant access to the same set, so the relevant screening dimension is quantity rather than model access. The different depletion rates across models (e.g., Opus consuming resources $5\times$ faster than Sonnet) function as heterogeneous per-unit costs within a single compute-budget mechanism, analogous to the token-class costs $c_j$ in \autoref{prop:packages}.

\begin{figure}[ht]
\centering
\begin{tikzpicture}[
    tier/.style={
        rectangle,
        draw=black!60,
        fill=white,
        minimum width=3.8cm,
        minimum height=6.2cm,
        rounded corners=3pt,
        align=center
    },
    header/.style={
        rectangle,
        fill=blue!15,
        minimum width=3.6cm,
        minimum height=0.8cm,
        rounded corners=2pt,
        font=\bfseries
    },
    price/.style={
        font=\normalsize\bfseries
    },
    item/.style={
        font=\small,
        text width=3.2cm,
        align=left
    }
]
\node[tier] (pro) at (4.2,0) {};
\node[header] at (4.2,2.4) {Pro};
\node[price] at (4.2,1.6) {\$20/mo};
\node[item] at (4.2,0.3) {
    \textbullet~All models\\[2pt]
    \textbullet~$\sim$45 msgs/5hrs\\[2pt]
    \textbullet~5$\times$ free tier\\[2pt]
    \textbullet~Projects, tools
};
\node[tier] (max5) at (8.4,0) {};
\node[header] at (8.4,2.4) {Max 5x};
\node[price] at (8.4,1.6) {\$100/mo};
\node[item] at (8.4,0.1) {
    \textbullet~All models\\[2pt]
    \textbullet~$\sim$225 msgs/5hrs\\[2pt]
    \textbullet~15--35 hrs Opus/wk\\[2pt]
    \textbullet~Priority access
};
\node[tier] (max20) at (12.6,0) {};
\node[header] at (12.6,2.4) {Max 20x};
\node[price] at (12.6,1.6) {\$200/mo};
\node[item] at (12.6,0.1) {
    \textbullet~All models\\[2pt]
    \textbullet~$\sim$900 msgs/5hrs\\[2pt]
    \textbullet~24--40 hrs Opus/wk\\[2pt]
    \textbullet~Full Opus access
};
\draw[->, thick, blue!60] (4, -3.8) -- (12.8, -3.8);
\node[below, font=\small\itshape] at (8, -4.0) {Increasing token budget $X(\theta)$};

\end{tikzpicture}
\caption{Anthropic subscription tiers (January 2026). Model access is constant across paid tiers; differentiation occurs through usage allocations.}
\label{fig:claude-pricing}
\end{figure}

\subsection{OpenAI}
 Unlike Anthropic's tier structure, which differentiates mostly through constraints on usage, 
OpenAI's ChatGPT pricing differentiates through both quantity and  exclusive model access: higher tiers grant access to more capable models and relax usage constraints. This joint screening aligns with the multi-model monopolist of Section~5.2.

\autoref{fig:openai-pricing} summarizes OpenAI's  tier structure as of January 2026. The Free tier provides limited GPT-4o access with automatic downgrade to GPT-4o-mini when capacity is constrained. Plus (\$20/month) expands to approximately 80 GPT-4o messages per 3-hour window and adds reasoning models (o3, o4-mini) with a separate weekly limit of roughly 100 queries. Pro (\$200/month) removes most constraints and provides exclusive access to o1-pro, OpenAI's most compute-intensive reasoning model. Thus, both Anthropic and OpenAI converge on similar price points (\$20 standard, \$200 premium), but implement very different screening strategies.

\begin{figure}[ht]
\centering
\begin{tikzpicture}[
    tier/.style={
        rectangle,
        draw=black!60,
        fill=white,
        minimum width=4.5cm,
        minimum height=6.4cm,
        rounded corners=3pt,
        align=center
    },
    header/.style={
        rectangle,
        fill=green!15,
        minimum width=4.3cm,
        minimum height=0.8cm,
        rounded corners=2pt,
        font=\bfseries
    },
    price/.style={
        font=\normalsize\bfseries
    },
    item/.style={
        font=\small,
        text width=4.3cm,
        align=left
    }
]
\node[tier] (free) at (0,0) {};
\node[header] at (0,2.5) {Free};
\node[price] at (0,1.7) {\$0/mo};
\node[item] at (0,-0.3) {
    \textbullet~GPT-4o (limited)\\[2pt]
    \textbullet~Downgrades to mini\\[2pt]
    \textbullet~No reasoning models\\[2pt]
    \textbullet~Basic features
};
\node[tier] (plus) at (4.8,0) {};
\node[header] at (4.8,2.5) {Plus};
\node[price] at (4.8,1.7) {\$20/mo};
\node[item] at (4.8,-0.3) {
    \textbullet~80 GPT-4o msgs/3hrs\\[2pt]
    \textbullet~Access to o3, o4-mini\\[2pt]
    \textbullet~100 reasoning msgs/wk\\[2pt]
    \textbullet~Custom GPTs
};
\node[tier] (pro) at (9.6,0) {};
\node[header] at (9.6,2.5) {Pro};
\node[price] at (9.6,1.7) {\$200/mo};
\node[item] at (9.6,-0.3) {
    \textbullet~Unlimited GPT-4o, o3\\[2pt]
    \textbullet~Exclusive o1-pro access\\[2pt]
    \textbullet~Highest compute\\[2pt]
    \textbullet~Priority capacity
};
\draw[->, thick, green!60] (1.5, -4.0) -- (9.6, -4.0);
\node[below, font=\small\itshape] at (5.55, -4.2) {Increasing model capability and token budget};

\end{tikzpicture}
\caption{OpenAI ChatGPT subscription tiers (January 2026). Higher tiers grant access to more capable models and larger usage allocations.}
\label{fig:openai-pricing}
\end{figure}

In \autoref{prop:packages-multiple}, we showed that when models differ in cost curvature, the optimal mechanism assigns higher-capability models to higher buyer types. OpenAI's decision to reserve its o1-pro model for the highest tier is consistent with this prediction. In consumer subscriptions, the relevant cost differences across models may reflect inference-time compute intensity rather than user fine-tuning, but the qualitative implication of a monotone assignment with discrete jumps at upgrade thresholds is the same.

\subsection{Model Aggregators: GitHub and Quora}

A number of AI platforms, including  GitHub Copilot and Quora's Poe, do not develop their own models.  Instead, they aggregate models produced by others  (OpenAI, Anthropic, Google, Mistral, and others) and sell access to users through a single interface. A common contractual arrangement is one whereby  subscribers pay a monthly fee, receive a budget of platform-specific credits, and allocate those credits across the available models. Selecting a more capable model typically depletes the budget faster. 
We now describe Quora's and GitHub's platforms in detail and then relate their  pricing structures to the indirect implementation of the optimal mechanism discussed in \autoref{sec:two-part-tariff}.

\paragraph{Quora-Poe}
Poe offers subscribers access to over 100 AI models through a unified chat interface.  A subscriber pays a fixed monthly fee and receives an allocation of ``compute points,'' which serve as the platform's internal currency.  Each model is priced in points per message at a rate that reflects its inference cost.\footnote{For instance, a message to GPT-4o-mini costs 9 points, while a message to Claude Opus~4 costs 4{,}105 points, a ratio of roughly $450\times$. }

Table~\ref{tab:poe-pricing} summarizes Poe's tier structure.  The entry-level Basic plan provides 300{,}000 points per month for \$5; the top-tier Premium plan provides 12{,}500{,}000 points for \$250.  Above the Basic tier, the effective price per million points is constant at \$20, so that higher tiers simply offer a proportionally larger budget at a fixed unit rate.  Table~\ref{tab:poe-models} reports point costs for selected models, illustrating the wide dispersion across model classes.

\begin{table}[ht]
\centering
\begin{tabular}{lrrr}
\toprule
Tier & Monthly Price & Points/Month & Eff.\ \$/M Points \\
\midrule
Basic & \$5 & 300,000 & \$16.67 \\
Standard & \$20 & 1,000,000 & \$20.00 \\
Pro & \$50 & 2,500,000 & \$20.00 \\
Advanced & \$100 & 5,000,000 & \$20.00 \\
Premium & \$250 & 12,500,000 & \$20.00 \\
\bottomrule
\end{tabular}
\caption{Poe subscription tiers (2025).}
\label{tab:poe-pricing}
\end{table}

\begin{table}[ht]
\centering
\begin{tabular}{llr}
\toprule
Class & Model & Points/Message \\
\midrule
Budget & GPT-4o-mini & 9 \\
 & Gemini-2.0-Flash & 9 \\
\midrule
Mid-tier & GPT-4o & 224 \\
 & Claude-3.5-Sonnet & 276 \\
\midrule
Frontier & Claude-3-Opus & 1,697 \\
 & Claude-Opus-4 & 4,105 \\
\bottomrule
\end{tabular}
\caption{Poe per-model point costs (2025).}
\label{tab:poe-models}
\end{table}

The connection to the maximum-spend framework of \autoref{sec:two-part-tariff} is immediate.  The monthly subscription fee corresponds to the transfer~$T_n$; the point allocation corresponds to the budget~$B_n$; and the per-model point costs correspond to marginal-cost pricing of different token classes.  The subscriber chooses how to allocate the budget taking the marginal prices into account. Finally, once the budget is exhausted, access to all models is suspended until the next billing cycle, and points do not roll over. Thus, Poe's menu is best captured by a maximum-spend mechanism.

\paragraph{GitHub Copilot}
GitHub Copilot provides AI-assisted coding within a developer's integrated  environment.  Like Poe, Copilot offers multiple pricing tiers, each bundling a monthly fee with a budget of ``premium requests'' that the developer allocates across model calls.  Table~\ref{tab:copilot-pricing} summarizes the plan structure.

\begin{table}[ht]
\centering
\begin{tabular}{lrrl}
\toprule
Plan & Monthly Fee & Premium Requests & Notes \\
\midrule
Free & \$0 & 50 & Limited quota, basic completions \\
Pro & \$10 & 300 & Unlimited completions, chat, IDE support \\
Pro+ & \$39 & 1{,}500 & Larger quota, advanced models \\
Business & \$19/user & 300/user & Org controls, policy management \\
Enterprise & \$39/user & 1{,}000/user & Enterprise features \\
\bottomrule
\end{tabular}
\caption{GitHub Copilot pricing tiers and premium request budgets (2026).}
\label{tab:copilot-pricing}
\end{table}

The mechanism by which Copilot meters usage differs slightly from Poe's point system but serves the same  function.  Rather than assigning each model a point cost per message, Copilot assigns each model a multiplier that determines how many premium requests a single interaction consumes.  A set of baseline models, including GPT-5 mini, GPT-4.1, and GPT-4o, carry a multiplier of~$0$ on paid plans: interactions with these models are included at no additional budget cost.  Efficient reasoning models such as Claude Haiku~4.5 and o4-mini carry multipliers of $0.25$--$0.33$.  Standard models such as Claude Sonnet~4.x and GPT-5.x  carry a multiplier of~$1$.  At the frontier, Claude Opus~4.5 carries a multiplier of~$3$, and GPT-4.5 carries a multiplier of~$50$.

 This mechanism matches  the framework of \autoref{sec:two-part-tariff} quite closely.  The plan fee corresponds to~$T_n$; the monthly premium-request allocation corresponds to~$B_n$; and the model multipliers play the role of heterogeneous marginal costs. 
A distinctive feature of Copilot is that it extends this budget mechanism with an overage option.  When a developer (i.e., a buyer) exhausts her monthly allocation,  Copilot allows continued use at a fixed charge of \$0.04 per premium request. The minimum-spend framework of \autoref{sec:two-part-tariff} matches this feature rather well: each user commits to a certain amount of monthly spending, there are no refunds, but the developer can increase her consumption ex post at linear prices.\footnote{As in the theory, the overage charge interacts with the multiplier system: a query to a $1\times$ model costs \$0.04 in the overage region, while a query to a $3\times$ model (Claude Opus~4.5) costs \$0.12.} 

Thus, both Poe and Copilot implement the core logic of the committed-spend mechanisms: a non-bankable budget, priced up front via a subscription fee, that the user freely allocates across inputs with heterogeneous per-unit costs.  The key distinction lies in how each platform handles the overages.  Poe enforces a hard cap: once points are exhausted, access is suspended until the next billing cycle.  This corresponds exactly to the maximum-spend mechanism.  Copilot, by contrast, allows the cap  to be relaxed at a known marginal price (which varies across models through the multiplier system), effectively implementing a minimum-spend mechanism.

\subsection{API Pricing}

Alongside consumer subscriptions, every major LLM provider 
operates a developer-facing API through which applications can 
programmatically submit prompts and receive completions. A 
developer who calls the API pays per token, with separate rates 
for input tokens (the prompt) and output tokens (the model's 
response). There is no subscription fee, no bundled allocation, 
and no minimum commitment: the developer pays only for what she 
uses. Table~\ref{tab:api-pricing} reports prices for selected 
models across three providers as of January~2025.

\begin{table}[ht]
\centering
\begin{tabular}{llrr}
\toprule
Provider & Model & Input (\$/M) & Output (\$/M) \\
\midrule
OpenAI & GPT-4o-mini & 0.15 & 0.60 \\
OpenAI & GPT-4o & 2.50 & 10.00 \\
OpenAI & o1 (reasoning) & 15.00 & 60.00 \\
\midrule
Anthropic & Claude 3.5 Haiku & 0.80 & 4.00 \\
Anthropic & Claude 3.5 Sonnet & 3.00 & 15.00 \\
Anthropic & Claude 3 Opus & 15.00 & 75.00 \\
\midrule
Google & Gemini 1.5 Flash & 0.075 & 0.30 \\
Google & Gemini 1.5 Pro & 1.25 & 5.00 \\
\bottomrule
\end{tabular}
\caption{API pricing across major providers (January 2025). Prices per million tokens.}
\label{tab:api-pricing}
\end{table}

The pricing structure is notably uniform. Output tokens are 
priced at $3$--$5\times$ input tokens across all providers, 
reflecting that output generation is sequential and 
autoregressive (each token requires a forward pass conditioned on
all preceding tokens), whereas input encoding is parallelizable 
and therefore substantially cheaper per token. Pricing is strictly 
linear: there are no quantity discounts, volume commitments, or 
tiered rates in standard API access. Although outside our model, 
all providers offer batch discounts (typically 50\%) for 
asynchronous processing and prompt-caching discounts (up to 90\%) 
for repeated inputs.

This linear, pay-per-token structure stands in sharp contrast to 
the nonlinear menus that characterize consumer subscriptions. It 
corresponds instead to the constrained-efficient allocation of 
\autoref{cor:capacity-efficient}: when a provider faces capacity 
constraints but wishes to maximize total surplus rather than 
profit (for instance, to lock in consumers in anticipation of future monetization), the 
constrained-efficient allocation can be implemented via linear 
prices equal to marginal costs inflated by shadow costs. The absence of nonlinear screening 
in API pricing is consistent with providers currently prioritizing 
adoption over rent extraction.

The uniformity of the pricing format across providers is itself evidence of competitive pressure: if any single provider were to introduce nonlinear screening in the API segment, it would risk losing developers to competitors who maintain simpler, linear pricing.

Evidence from pricing dynamics is consistent with this interpretation. First, API prices have declined rapidly: GPT-4-class capability fell from \$30/\$60 per million tokens at launch (March 2023) to \$2.50/\$10 by August 2024, a 90\% reduction in 16 months. Second, providers appear to cut prices in response to competitive launches rather than cost improvements; OpenAI's August 2024 reductions followed Claude~3.5~Sonnet and Llama~3.1 releases. Third, gross margins of 50--75\% imply prices remain above marginal cost but are being compressed by competition. The entry of DeepSeek in January 2025 with reasoning-model pricing at \$0.55/\$2.19, versus OpenAI's \$15/\$60 for comparable capability, suggested that further compression is viable. See \cite{defr25} for a comprehensive treatment of market share dynamics in the API segment.

\section{Conclusion}

A priori, the problem of pricing LLM access appears intractable: users have high-dimensional private information, the allocation space is also high-dimensional, and the user's hidden allocation of tokens across tasks introduces moral hazard. In this paper, we have shown that homogeneity of the gain function generates a sufficient-statistic reduction that makes the problem solvable. Users' high-dimensional type profiles collapse to a scalar aggregate type; the seller's problem reduces to one-dimensional screening; and the optimal mechanism admits simple implementations (committed-spend contracts and two-part tariffs) that correspond precisely to the pricing structures observed at leading providers.\vspace{.25in}

While we have phrased the analysis in terms of large language models, the framework applies more broadly to any setting in which a provider sells access to a general-purpose technology with multiple input classes and heterogeneous users. Cloud computing, where providers offer a multiplicity of services with both horizontal and vertical differentiation, is a natural example. In both settings, the core pricing problem is how to allocate and monetize costly computing resources, particularly when operating under capacity constraints.

Several extensions would enrich the analysis.  First, we assumed that all tasks are homogeneous in their use of input, output, and fine-tuning tokens, and differ only vertically in how valuable a given task is to the buyer. Allowing the gain function parameters to vary across tasks would break the aggregation result. Understanding how the optimal mechanism changes when the scalar sufficient statistic no longer exists is a challenging but important open question, especially in the context of competing specialized models.

Second, we assumed that all buyer types have fine-tuning data readily available. In practice,  buyers face constraints on data availability or may be unwilling to share data with the provider because of privacy, compliance, or intellectual-property concerns. A data-poor buyer may then substitute toward inference-time tokens, while a data-rich buyer substitutes toward fine-tuning tokens. When data availability is the user's private information, the provider's problem involves screening on two dimensions (aggregate type and data availability), which may require new techniques.

Third, LLM platforms exhibit network effects and data externalities that our static framework abstracts from. A provider that attracts more users may improve its model through fine-tuning, reinforcement learning from human feedback, or simply from the revenue that funds further training. These dynamic complementarities between adoption and model quality could substantially reshape optimal pricing, potentially justifying the aggressive below-average-cost pricing observed in the API segment not just as a means to capture market share but as investment in model improvement.

Fourth, our analysis treats each provider as either a monopolist or a leader facing a competitive fringe. A full oligopoly analysis with multiple differentiated proprietary models would capture the strategic interactions among Anthropic, OpenAI, Google, and others that increasingly shape the market. The tractability of our framework through  the reduction to one-dimensional types suggests that such an extension may be feasible and would yield further predictions about equilibrium pricing and product differentiation.

The LLM industry is at an inflection point between growth-oriented pricing and profit-maximizing pricing. Our framework provides a theoretical foundation for understanding the mechanisms through which this transition will unfold, and for evaluating whether the resulting market structure efficiently allocates this critical economic input.

\pagebreak
\appendix
\section{Proofs and Derivations}\label{app:proofs}
\paragraph{Proof of \autoref{prop:efficient}}
For any $z$ and  $i$, the optimal allocation of inference tokens on task $i$ solves
\begin{equation*}
  \max_{x_{ij}\geq0} \ w_i\Psi(x_i)\Phi(z)-\sum_{j=1}^{J}c_jx_{ij}.
\end{equation*}
If $w_i=0$ or $\Phi(z)=0$, then $x_i=0$. Otherwise, by the Inada assumption, the problem admits an interior solution,  which satisfies the system of $J$ first-order conditions:
\begin{equation}\label{eq:eff-foc}
  w_i  \nabla\Psi (x_i)\Phi(z)=c. 
\end{equation}
By (\ref{eq:eff-foc}), for all $i$, $\nabla\Psi(x_i)$ belongs to a ray with a direction $c$. Because $\Psi$ is homogeneous and strictly concave, the ray in the space of gradients corresponds to a ray in the space of tokens.\footnote{By homogeneity, for all $x\in\reals^J_+$ and $\ind>0$, $\nabla\Psi(\ind x)=\ind^{\sigma-1}\nabla\Psi(x)$, so any token ray corresponds to a gradient ray. By strict concavity, if $x_1\neq x_2$, then $\nabla\Psi(x_1)\neq\nabla\Psi(x_2)$, so distinct token rays map to distinct gradient rays.
} Thus, for each $i$, any optimal $x_i$ can be written as:
\begin{equation*}
    x_i=d y_i,
\end{equation*}
where $d\in\reals^J_+$ is the unique vector that solves $\nabla\Psi(d)=c$. 

Therefore, by the homogeneity of $\Psi$, we have $\nabla \Psi (x_i)=y_i^{\sigma-1}\nabla \Psi (d)$, and (\ref{eq:eff-foc}) can be rewritten as:
\begin{equation*}
    w_i y_i^{\sigma-1}\nabla\Psi(d)\Phi(z)=\nabla\Psi(d),
\end{equation*}
which pins down the optimal scale as:
\begin{equation*}
      y_i=w_i^{\frac{1}{1-\sigma}}\Phi(z)^{\frac{1}{1-\sigma}}.
\end{equation*}

The resulting total surplus ignoring fine-tuning costs is:
\begin{align}
  \int_{0}^{1} w_i \Psi(x_i) \Phi(z) di-\sum_{j=1}^{J}c_j\int_{0}^{1}x_{ij} di&=\int_0^1 w_i y_i^\sigma \Psi(d) \Phi(z) -y_i (d\cdot c)\, di \notag\\ 
  =\int_0^1 w_i^{\frac{1}{1-\sigma}}\Phi(z)^{\frac{1}{1-\sigma}}(\Psi(d)-d\cdot c)\, di  
 &=\theta(\w)^{\frac{1}{1-\sigma}}\Phi(z)^{\frac{1}{1-\sigma}} (1-\sigma)\Psi(d),\label{eq:surplus-net-of-fine-tuning}
\end{align}
where $\theta(\w)$ is the \emph{aggregate type} defined as \begin{equation}
    \theta(\w)\triangleq \left(\int_0^1 w_i^{\frac{1}{1-\sigma}} di\right)^{1-\sigma},
\end{equation}
and the last equality in (\ref{eq:surplus-net-of-fine-tuning}) follows from the definition of $d$ and  Euler's theorem for homogeneous functions: $d\cdot\nabla\Psi(d)=\sigma\Psi(d)$.

Similarly, the total amount of inference tokens of class $j$ is:
\begin{equation*}
    \int_0^1 x_{ij} di=\int_0^1 d_j y_i di=\int_0^1  d_j w_i^{\frac{1}{1-\sigma}}\Phi(z)^{\frac{1}{1-\sigma}} di=\theta(\w)^{\frac{1}{1-\sigma}} d_j \Phi(z)^{\frac{1}{1-\sigma}}.
\end{equation*}
This completes the proof.\hfill$\square$

\paragraph{Proof of \autoref{cor:capacity-efficient}}

The Lagrangian approach applies. Thus, associating Lagrange multipliers $\lambda_j\geq0$ and $\hat\lambda_k\geq0$ with the capacity constraints for the different token classes, the optimal solution solves
\begin{equation*}
  \max_{(x_i)_{i\in[0,1]},z\geq0}\int_{0}^{1} w_i g(x_i,z)di-\sum_{j=1}^{J}(c_j+\lambda_j)\int_{0}^{1}x_{ij}di-\sum_{k=1}^{K}(\hat c_k +\hat \lambda_k) z_k.
\end{equation*}
As such, the solution is an efficient allocation given the adjusted costs $c_j'\triangleq c_j+\lambda_j$ and $\hat c_k'\triangleq \hat c_k+\hat \lambda_k$. The result follows.\hfill$\square$

\paragraph{Proof of \autoref{lem:buyer-optimal-package}}

Consider the buyer-optimal inference token allocation across tasks for any given token budgets $(X,Z)$ such that $\Phi(Z)>0$:
\begin{equation}\label{eq:token-budget-buyer-problem}
    \max_{x_{ij}\geq0} \int_0^1 w_i \Psi(x_i)\Phi(Z)di,\quad\mathrm{s.t.}\ \int_0^1 x_{ij}di=X_j \textrm{ for } j\in[J].
\end{equation}
Because $\Phi(Z)$ is a strictly positive constant, it can be ignored. Assume that for each $j$, $X_j>0$; otherwise, $x_{ij}=0$ for all $i$. Lagrangian approach applies. We associate Lagrange multipliers $\lambda=(\lambda_1,\dots,\lambda_J)$ with the budget constraints of (\ref{eq:token-budget-buyer-problem}); the resulting first-order conditions are:
\begin{equation*}
    w_i\nabla\Psi(x_i)=\lambda.
\end{equation*}
By the same logic as in the efficiency analysis, these conditions imply that the relative ratios of input tokens are the same across all tasks:
\begin{equation*}
    x_{i}= y_i d,
\end{equation*}
for some $d\in\reals^J_+$. The budget constraints imply that $(\int_0^1 y_i di) d=X$ and hence $d$ is proportional to $X$ and can be normalized to satisfy $\sum_{j=1}^J d_j=1$, leading to
\begin{equation}\label{eq:budget-d}
    d =\frac{1}{\sum_{j=1}^J X_j} X.
\end{equation}
The optimal scales $y_i$ solve:
\begin{equation*}
    \max_{y_i\geq 0} \int_0^1 w_i y_i^{\sigma} \Psi(d) di,\quad\mathrm{s.t.}\  \int_0^1 y_i di=\sum_{j=1}^J X_j.
\end{equation*}
At the optimum, the marginal gains  $w_i\sigma y_i^{\sigma-1}$ are equalized across tasks, so:
\begin{equation*}
    y_i=w_i^{\frac{1}{1-\sigma}}\frac{\sum_{j=1}^J X_j}{\int_0^1 w_{\ind}^{\frac{1}{1-\sigma}}d\ind}.
\end{equation*}
The optimal buyer's payoff is:
\begin{align*}
    U&=\int_0^1 w_i \left(w_i^{\frac{1}{1-\sigma}}\frac{\sum_{j=1}^J X_j}{\int_0^1 w_{\ind}^{\frac{1}{1-\sigma}}d\ind}\right)^{\sigma}\Psi(d)\Phi(Z)di\\
    &=\left(\int_0^1 w_i^{\frac{1}{1-\sigma}}di\right)^{1-\sigma}\left(\sum_{j=1}^J X_j\right)^\sigma\Psi(d)\Phi(Z)=\theta(\w)\Psi(X)\Phi(Z),
\end{align*}
where in the last equality we used the homogeneity of $\Psi$ and (\ref{eq:budget-d}).\hfill\qed

\paragraph{Proof of \autoref{lem:cost-function}}

The fact that $C(Q)$ is strictly increasing and strictly convex is immediate because $g(X,Z)=\Psi(X)\Phi(Z)$ is strictly increasing and strictly concave in its arguments whenever $g(X,Z)>0$.

To show $C'_+(0)=0$, observe that since $C(Q)$ is convex, $C'_+(0)=\inf_{Q>0} C(Q)/Q$. Pick any $X_0$ such that $\Psi(X_0)>0$ and $Z_0$ as in Footnote \ref{foot:inada}. For $\ind>0$, consider $(X_\ind,Z_\ind)=\ind(X_0,Z_0)$ and set $Q_\ind=g(X_\ind,Z_\ind)$. We have,
\begin{equation*}
    \frac{C(Q_\ind)}{Q_\ind}\leq \frac{c\cdot X_\ind+\hat c\cdot Z_\ind}{Q_\ind}=\frac{\ind(c\cdot X_0+\hat c\cdot Z_0)}{\ind^\sigma\Psi(X_0)\Phi(\ind Z_0)}=\frac{c\cdot X_0+\hat c\cdot Z_0}{\Psi(X_0)}\frac{1}{\Phi(\ind Z_0)/\ind^{1-\sigma}}.
\end{equation*}
By the choice of $Z_0$, $\Phi(\ind Z_0)/\ind^{1-\sigma}\to+\infty$ as $\ind\downarrow 0$, hence the right-hand side goes to $0$. Since $Q_\ind\downarrow 0$, it follows that $\inf_{Q>0} C(Q)/Q=0$.\hfill$\square$

\paragraph{Proof of \autoref{prop:packages}} The result is an application of \cite{mussa1978monopoly} to the buyer's payoff $\theta Q$ from \autoref{lem:buyer-optimal-package} and the cost function $C(Q)$ from \autoref{lem:cost-function}. The optimal quality schedule (\ref{eq:quality_package}) and transfers (\ref{eq:transfers_package}) follow from standard arguments.\hfill$\square$

\paragraph{Proof of \autoref{prop:indirect-implementation}} Fix an optimal direct mechanism $(X^*(\theta),Z^*(\theta),T^*(\theta))$ with the associated aggregate quality $Q^*(\theta)=\Psi(X^*(\theta))\Phi(Z^*(\theta))$.

\textbf{Maximum-spend mechanism.} Consider a maximum-spend mechanism with $p=(c,\hat c)$, $T(\theta)=T^*(\theta)$, and $B(\theta)=C(Q^*(\theta))$. Under this mechanism, since tokens are priced at marginal cost, each type $\w$ purchasing budget $B(\theta')$ would optimally allocate tokens in a constrained-efficient way, thus deriving payoff $\theta(\w) Q$ where $C(Q)=B(\theta')$, i.e., $Q=Q^*(\theta')$. Hence, the choice across items in this mechanism is equivalent to the choice in the direct mechanism, and this mechanism implements the allocation of the direct mechanism.

\textbf{Two-part-tariff mechanism.}  When type $\w$ faces any two-part tariff, his optimal token allocation problem is equivalent to the efficient allocation problem (\ref{eq:planner-problem}) with prices taking the roles of costs. Therefore, all types $\w$ with the same $\theta(\w)$ have the same payoff from any possible two-part tariff and can be treated as a single type.

Faced with a given item, the problem of type $\theta$ can be written as value-maximization-payment-minimization:
\begin{equation*}
    \max_{Q\geq0} \ \theta Q - P(Q),
\end{equation*}
where
\begin{equation*}
    P(Q)\triangleq\min_{X_1,\dots,X_J, Z_1,\dots,Z_K\geq0} \sum_{j=1}^J p_j X_j+\sum_{k=1}^K \hat p_k Z_k,\quad \mathrm{s.t. }\ \Psi(X)\Phi(Z)=Q.
\end{equation*}

Consider the following two-part-tariff mechanism:
\begin{align*}
       p_j(\theta)&=m(\theta)c_j,\ \hat p_k(\theta)=m(\theta)\hat c_k,\  j\in[J],\ k\in[K],\\
       p_0(\theta)&=t(\theta)-m(\theta)C(Q(\theta)),
\end{align*}
where   $C(Q)$ is as defined in (\ref{eq:costs_package}), $Q(\theta)$ is as defined in (\ref{eq:quality_package}), and $t(\theta)$ is as defined in (\ref{eq:transfers_package}). Under this mechanism, $\sum_{j=1}^J p_j X_j+\sum_{k=1}^K \hat p_k Z_k=m(\theta)(\sum_{j=1}^J c_j X_j+\sum_{k=1}^K \hat c_k Z_k)$. Thus, the buyer-optimal allocation  is efficient, and $P(Q)=m(\theta)C(Q)$.

The rest of the argument is standard (e.g., \citet[pp. 154-157]{tiro88}). A buyer-optimal $Q(\theta)$ is determined by the first-order condition:
\begin{equation*}
    \theta=P'(Q(\theta))=m(\theta)C'(Q(\theta)),
\end{equation*}
and thus the buyer-optimal level of quality satisfies the  optimality condition:
\begin{equation*}
    \MRI(\theta)=C'(Q(\theta)).
\end{equation*}
If $m(\theta)$ is decreasing, then the implied payment schedule in units of $C(Q)$ is concave.
Thus, a menu of two-part tariffs with markups $m(\theta)$, constant across inputs, implements the desired quality schedule $Q(\theta)$, with each type $\theta$ consuming the optimal amount of tokens and paying the optimal total transfer.

\textbf{Minimum-spend mechanism.} Consider a minimum-spend mechanism that consists of $p(\theta)=r(\theta)(c,\hat c)$ and $B(\theta)=T^*(\theta)$ for all  types such that $Q^*(\theta)>0$, where $r(\theta)=\frac{T^*(\theta)}{C(Q^*(\theta))}$. By construction, since prices are proportional to costs, type $\w$, when consuming aggregate quality $Q$, would optimally derive value $\theta(\w) Q$. Thus, the buyer's problem of optimal spending can be formulated in terms of $Q$. The restriction of minimum spend when reporting $\theta'$ can be written as $r(\theta')C(Q)\geq T^*(\theta')$, which simplifies to $Q\geq Q^*(\theta')$. Thus, the optimal payoff of type $\theta$ when reporting type $\theta'$ is
\begin{equation*}
U(\theta,\theta')=\max_{Q\geq Q^*(\theta')} \theta Q-r(\theta')C(Q).
\end{equation*}

This mechanism implements the allocation of the direct mechanism if each type prefers to report truthfully and choose $Q=Q^*(\theta)$. Since $B(\theta)$ is increasing, for this to happen, $r(\theta)$ must be decreasing.

In fact, $r(\theta)$ being decreasing is not only necessary but also sufficient for implementability. To see this, note that $Q^*(\theta)$ is continuously increasing and $r(\theta)\geq 1$. Thus, the optimal deviation is always weakly below the maximum quality offered in the direct menu:
\begin{equation*}
\arg\max_{Q\geq Q^*(\theta')} \theta Q-r(\theta')C(Q)\leq \arg\max_{Q\geq Q^*(\theta')} \overline \theta Q- C(Q)=Q^{*}(\overline \theta),
\end{equation*}
where $\overline \theta$ is the maximum point in $\mathrm{supp}\,F$. Moreover, if type $\theta$ deviates to type $\theta'$ and consumes quality $Q\leq Q^*(\overline \theta)$, then there exists a type $\theta''\geq\theta'$ such that $Q=Q^*(\theta'')$. Since $r$ is decreasing, it is more profitable for type $\theta$ to deviate to type $\theta''$ and consume $Q^*(\theta'')$. But such a deviation is equivalent to the deviation to $\theta''$ in the direct menu and is therefore not profitable.

It remains to show that decreasing $m$ implies decreasing $r$. Assume that $m$ is decreasing. Since $m$ and $r$ are continuous, it suffices to show that their derivatives are negative at all differentiable points. At those points, $m'(\theta)\leq0$, which implies $\MRI(\theta)-\MRI'(\theta)\theta\leq 0$, and
\begin{equation*}
r'(\theta)=\frac{T^{*\prime}(\theta)C(Q^*(\theta))-T^*(\theta)C'(Q^*(\theta))Q^{*\prime}(\theta)}{C(Q^*(\theta))^2}=\frac{Q'(\theta)}{C(Q^*(\theta))^2}(\theta C(Q^*(\theta))-T^*(\theta)\MRI(\theta)),
\end{equation*}
where we used the fact that $T^{*\prime}(\theta)=\theta Q^{*\prime}(\theta)$ and $C'(Q^*(\theta))=\MRI(\theta)$. Since $Q$ is increasing, it suffices to show that $\frac{\theta}{\MRI(\theta)}\leq \frac{T^*(\theta)}{C(Q^*(\theta))}$, i.e., the markup in the two-part tariff implementation is lower than the markup in the minimum-spend implementation. To show this, observe that at the maximal excluded type $\theta_0$, $\theta C(Q^*(\theta))-T^*(\theta)\MRI(\theta)=0$, and for all types $\theta>\theta_0$,
\begin{equation}
\left( \frac{\theta  C(Q^*(\theta))-T^*(\theta)\MRI(\theta)}{\theta}\right)'=\frac{T(\theta)}{\theta^2}(\MRI(\theta)-\MRI'(\theta)\theta)\leq 0.
\end{equation}
Therefore, $\theta C(Q^*(\theta))-T^*(\theta)\MRI(\theta)\leq 0$, and $r$ is decreasing.\hfill\qed

\paragraph{Proof of \autoref{lem:buyer-optimal-multiple}}
For general, not necessarily identical, $\sigma_l$, the marginal gain from assigning an infinitesimal task with value $w_i$ to model $l$ is,  by (\ref{eq:buyer-optimal-package-l}):
\begin{equation}
   \delta_l(w_i)=\frac{1-\sigma_l}{(\int_{I_l} w_\ind^{1/(1-\sigma_l)} d\ind)^{\sigma_l}}Q_l w_i^{1/(1-\sigma_l)}.
\end{equation}
Under an optimal task split, if a task with value $w_i$ is assigned to a model $l$, then it must be that for all $m$, $\delta_l(w_i)\geq \delta_m(w_i)$. Because $\delta_m(w_i)/\delta_l(w_i)$ is increasing in $w_i$ for all $m>l$, it follows that an optimal task-model allocation exists that forms a monotone partition, with models of higher $l$ (and thus higher $\sigma_l$) being allocated the more valuable tasks.

Denoting by $i_l$ the delimiting points, with $i_0=0$ and $i_L=1$, the optimal  partition is determined by equalizing the marginal gains, $\delta_l(w_{i_l})=\delta_{l+1}(w_{i_l})$ for $l\in[L-1]$:
\begin{equation}\label{eq:interval_split}
    \frac{(1-\sigma_l) Q_l w_{i_l}^{1/(1-\sigma_l)}}{(1-\sigma_{l+1})Q_{l+1}w_{i_l}^{1/(1-\sigma_{l+1})}}=\frac{(\int_{I_l} w_i^{1/(1-\sigma_l)} di)^{\sigma_l}}{(\int_{I_{l+1}} w_i^{1/(1-\sigma_{l+1})} di)^{\sigma_{l+1}}}.
\end{equation}
When $\sigma_l\equiv\sigma$, the optimality condition (\ref{eq:interval_split}) simplifies to the following expression:\footnote{In the general case of heterogeneous $\sigma_l$, the optimal task split and the resulting buyer's payoff do not admit tractable closed-form solutions. In particular, the fine details of the profile $\w$ could matter.}
\begin{equation*}
    \frac{ Q_l}{Q_{l+1}}=\frac{(\int_{I_l^*} w_i^{1/(1-\sigma)} di)^{\sigma}}{(\int_{I_{l+1}^*} w_i^{1/(1-\sigma)} di)^{\sigma}}.
\end{equation*}
Thus,
\begin{equation*}
    \int_{I^*_l} w_i^{1/(1-\sigma)} di=\frac{Q_l^{1/\sigma}}{\sum_m Q_m^{1/\sigma}}\int_0^1 w_i^{1/(1-\sigma)}di,
\end{equation*}
or, equivalently,
\begin{equation*}
    \theta_l(I^*_l)=\left(\frac{Q_l^{1/\sigma}}{\sum_m Q_m^{1/\sigma}}\right)^{1-\sigma}\theta.
\end{equation*}
The collection $(\theta_l(I^*_l))_{l=1}^L$ determines an optimal (monotone) partition $(I^*_l)_{l=1}^L$.\footnote{Non-monotone partitions may also be optimal, but $\theta_l(I^*_l)$ are uniquely determined.} All models with $Q_l>0$ are  employed at some tasks. The resulting optimal payoff is then (\ref{eq:buyer-value-multimodel}).\hfill$\square$

\paragraph{Proof of \autoref{lem:homog-indirect-cost}}
Drop the $l$ index. First, consider the case $\hat \sigma>0$. Define the unit-cost indices
\begin{equation*}
e  \triangleq \min_{x\ge 0}\{ c\cdot x:  \Psi(x)\ge 1 \},
\quad
\hat e  \triangleq \min_{z\ge 0}\{ \hat c\cdot z: \Phi(z)\ge 1 \}.
\end{equation*}
For strictly concave, increasing, homogeneous $\Psi,\Phi$ these are finite and attained, and the respective minimizers $d,\hat d$ are unique.

By homogeneity and the definition of $e$, the minimal cost to reach $\Psi(x)=\Psi_0$ is
\[
\min_{x\ge 0}\{c\cdot x:\Psi(x)\ge \Psi_0\}=\Psi_0^{1/\sigma}e,
\]
attained at $x=\Psi_0^{1/\sigma} d$. Similarly,
$\min_{z\ge 0}\{\hat c\cdot z:\Phi(z)\ge \Phi_0\}=\Phi_0^{1/\hat\sigma}\hat e$, attained at $z=\Phi_0^{1/\hat\sigma} \hat d$.

The cost minimization reduces to
\begin{equation*}
\min_{\Psi_0,\Phi_0\geq 0}\Psi_0^{1/\sigma}e+\Phi_0^{1/\hat\sigma}\hat e\quad\mathrm{s.t.}\quad \Psi_0 \Phi_0=Q.
\end{equation*}
Straightforward calculation gives:
\begin{equation*}
\Psi_0=Q^{\frac{\sigma}{\sigma+\hat\sigma}}\left(\frac{\hat e\,\sigma}{e\,\hat\sigma}\right)^{\frac{\sigma\hat\sigma}{\sigma+\hat\sigma}},\quad
\Phi_0=Q^{\frac{\hat\sigma}{\sigma+\hat\sigma}}\left(\frac{\hat e\,\sigma}{e\,\hat\sigma}\right)^{-\frac{\sigma\hat\sigma}{\sigma+\hat\sigma}},
\end{equation*}
which corresponds to the optimal token allocation:
\begin{equation*}
x^*(Q)=Q^{\frac{1}{\sigma+\hat\sigma}}
\Bigl(\frac{\hat e\,\sigma}{e\,\hat\sigma}\Bigr)^{\frac{\hat\sigma}{\sigma+\hat\sigma}} d,\quad
z^*(Q)=Q^{\frac{1}{\sigma+\hat\sigma}}
\Bigl(\frac{\hat e\,\sigma}{e\,\hat\sigma}\Bigr)^{-\frac{\sigma}{\sigma+\hat\sigma}}\hat d.
\end{equation*}
The resulting cost function is:
\begin{equation*}
C(Q)=e \Psi_0^{1/\sigma}+\hat e \Phi_0^{1/\hat\sigma}
=e^{\frac{\sigma}{\sigma+\hat\sigma}} \hat e^{\frac{\hat\sigma}{\sigma+\hat\sigma}}
\Bigl[\Bigl(\frac{\sigma}{\hat\sigma}\Bigr)^{\frac{\hat\sigma}{\sigma+\hat\sigma}}
+\Bigl(\frac{\hat\sigma}{\sigma}\Bigr)^{\frac{\sigma}{\sigma+\hat\sigma}}\Bigr]Q^{\frac{1}{\sigma+\hat\sigma}}=C_0 Q^{\frac{1}{\sigma+\hat\sigma}}.
\end{equation*}

The case $\hat \sigma=0$ is analogous and simpler because in that case $\Phi(z)\equiv \Phi_0>0$ and fine-tuning can be ignored. The resulting cost function is
\begin{equation*}
C(Q)=e \left(\frac{Q}{\Phi_0}\right)^{\frac{1}{\sigma}}=C_0 Q^{\frac{1}{\sigma}}.
\end{equation*}
This completes the proof.\hfill$\square$

\paragraph{Proof of \autoref{prop:efficient-multiple}} By \autoref{lem:homog-indirect-cost},  each model's surplus $\theta Q - \kappa_l Q^{1/(\sigma+\hat{\sigma}_l)}$ is concave in $Q$ and can be maximized in closed form. The efficient allocation selects the model that yields the highest surplus, yielding (\ref{eq:l-eff-multimodel})--(\ref{eq:q-eff-multimodel}). Single-model usage follows because the cost function (\ref{eq:cost-multimodel}) is concave in $Q^{1/\sigma}$.\hfill$\square$

\paragraph{Proof of \autoref{prop:packages-multiple}} The result follows by pointwise maximization of virtual surplus with $\MR(\theta)$ in place of $\theta$ in \autoref{prop:efficient-multiple}. Monotonicity of $\MR$ ensures that the pointwise solution $Q^{\mathrm{m}}(\theta)=Q^{*}(\MR(\theta))$ is increasing and therefore implementable.\hfill$\square$
\paragraph{Proof of \autoref{lemma:buyer_pay}}
When $\hat{\sigma}_F=0$,
   (\ref{eq:fringe-foc-1}) simplifies into:
\begin{equation}\label{eq:fringe-foc-2}
   q_L^{(\sigma+\hat{\sigma}_L)/\sigma}+q_F=\left(\frac{\theta \sigma}{c_F}\right)^{1/(1-\sigma)}.
\end{equation}
The outside option corresponding to $q_L=0$ is
\begin{equation}
    \psi(\theta)=\theta^{\frac{1}{1-\sigma}}(1-\sigma)\left(\frac{\sigma}{c_F}\right)^{\sigma/(1-\sigma)}.
\end{equation}
The formulation (\ref{eq:payoff_norm}) follows.\hfill\qed

\paragraph{Proof of \autoref{prop:fringe}} 
Denote the expression in the brackets of (\ref{eq:virtual-surplus-fringe}) by $\Pi(\theta,q)$ and observe that 
\begin{align}
  \Pi(\theta,q(\theta))=\begin{cases}
c_F q(\theta)^{(\sigma+\hat{\sigma}_L)/\sigma}-c_Lq(\theta), & \textrm{if\ }q(\theta)<\qh(\theta),\\
 \MR(\theta)q(\theta)^{\sigma+\hat{\sigma}_L}-c_Lq(\theta)-\psi(\theta)+\frac{1-F(\theta)}{f(\theta)}\psi'(\theta),& \textrm{if\ }q(\theta)>\qh(\theta). 
\end{cases} 
\end{align}
Consider the pointwise maximization of $\Pi(\theta,q(\theta))$. In the region $q(\theta)\in[0,\qh(\theta)$], $\Pi(\theta,q(\theta))$ is convex in $q(\theta)$ and therefore attains its  maximum  at a corner, $q(\theta)=0$ or $q(\theta)=\qh(\theta)$. Direct calculation shows that $\Pi(\theta,0)>\Pi(\theta,\qh(\theta))$ if and only if $\theta<\theta_1$, where
\begin{equation}
    \theta_1\triangleq \frac{c_F}{\sigma} \left(\frac{c_L}{c_F}\right)^{(1-\sigma)(\sigma+\hat{\sigma}_L)/\hat{\sigma}_L}.
\end{equation}

In the region $q(\theta)\geq\qh(\theta)$, if $\MR(\theta)\leq 0$, then optimally $q(\theta)=\qh(\theta)$. If $\MR(\theta)> 0$, then $\Pi(\theta,q(\theta))$ is concave in $q(\theta)$ and therefore the maximum is either at the corner, $q(\theta)=\qh(\theta)$, or in the interior, in which case it equals the optimal monopoly quantity
\begin{equation}
    q^{\mathrm{m}}(\theta)=\left(\frac{(\sigma+\hat{\sigma}_L)\MR(\theta)}{c_L}\right)^{1/(1-\sigma-\hat{\sigma}_L)}.
\end{equation}
Thus, the solution over the region $q(\theta)\geq\qh(\theta)$ is interior if and only if $\Pi_q(\theta,\qh(\theta))>0$, or
\begin{equation}\label{eq:cond_binding}
    \MR(\theta)>\frac{c_L}{\sigma+\hat{\sigma}_L}\left(\frac{\theta\sigma}{c_F}\right)^{\frac{\sigma(1-\sigma-\hat{\sigma}_L)}{(1-\sigma)(\sigma+\hat{\sigma}_L)}}.
\end{equation}
Condition (\ref{eq:cond_binding}) can hold on disjoint intervals of $\theta$ even if $\MR(\theta)$ is increasing, because the right-hand side is increasing in $\theta$. However, the power of $\theta$ on the right-hand side is strictly less than $1$. Thus, if $F$ satisfies monotone hazard rate, then $\MR'(\theta)\geq 1$, and condition (\ref{eq:cond_binding}) holds, if at all, for all $\theta>\theta_2$ where  $\theta_2$ solves:
\begin{equation}\label{eq:cond_binding_2}
    \MR(\theta_2)=\frac{c_L}{\sigma+\hat{\sigma}_L}\left(\frac{\theta_2\sigma}{c_F}\right)^{\frac{\sigma(1-\sigma-\hat{\sigma}_L)}{(1-\sigma)(\sigma+\hat{\sigma}_L)}}.
\end{equation}
If (\ref{eq:cond_binding_2}) doesn't admit a solution, then the solution is never interior and thus for all $\theta<\theta_1$, $q(\theta)=0$ and for all $\theta>\theta_1$, $q(\theta)=\qh(\theta)$. If (\ref{eq:cond_binding_2}) admits a solution at $\theta_2>\theta_1$, then: for all $\theta<\theta_1$, $q(\theta)=0$; for all $\theta\in(\theta_1,\theta_2)$,  $q(\theta)=\qh(\theta)$; for all $\theta\geq\theta_2$, $q(\theta)=q^{\mathrm{m}}(\theta)$. Finally, if (\ref{eq:cond_binding_2}) admits a solution at $\theta_2<\theta_1$, then  $\theta_3\in[\theta_2,\theta_1]$ exists such that for all $\theta<\theta_3$, $q(\theta)=0$ and for all $\theta\geq\theta_3$, $q(\theta)=q^{\mathrm{m}}(\theta)$. (The threshold $\theta_3>\theta_2$ is determined by the condition $\Pi(\theta_3,q^{\mathrm{m}}(\theta_3))=0$.)

Since all these (relaxed) solutions are implementable, the result follows.\hfill$\square$

\paragraph{Leader-Fringe Competition with Uniform Types} \strut

\noindent\emph{Competitive Mechanism---}
With $\hat{\sigma}_F=0$ the buyer's first-order condition implies
\[
q_L^{(\sigma+\hat{\sigma}_L)/\sigma} + q_F =
\Big(\tfrac{\theta\,\sigma}{c_F}\Big)^{\!1/(1-\sigma)}.
\]
On the \emph{fringe-indifference boundary} ($q_F=0$) the leader's quantity solves
\begin{equation*}
\hat q(\theta)
=
\Big(\tfrac{\theta\,\sigma}{c_F}\Big)^{\!\frac{\sigma}{(1-\sigma)(\sigma+\hat{\sigma}_L)}}.
\end{equation*}
When the buyer single-homes on the leader and the fringe constraint is slack,
\begin{equation*}
q^{\mathrm{m}}(\theta)
=
\Big(\tfrac{(\sigma+\hat{\sigma}_L)\,\MR(\theta)}{c_L}\Big)^{\!\frac{1}{1-\sigma-\hat{\sigma}_L}}
=
\Big(\tfrac{(\sigma+\hat{\sigma}_L)\,(2\theta-1)}{c_L}\Big)^{\!\frac{1}{1-\sigma-\hat{\sigma}_L}}.
\end{equation*}
 
We want  the optimal menu to be characterized by $0<\theta_1<\theta_2<1$ such that
\begin{align*}
q^{LF}(\theta)=
\begin{cases}
0, & \theta\le \theta_1,\\ 
\hat q(\theta), & \theta_1<\theta<\theta_2,\\ 
q^{\mathrm{m}}(\theta), & \theta\ge \theta_2,
\end{cases}
\qquad
q^{LF}_F(\theta)=
\begin{cases}
\Big(\dfrac{\theta\,\sigma}{c_F}\Big)^{\!\frac{1}{1-\sigma}}, & \theta\le \theta_1,\\ 
0, & \theta> \theta_1.
\end{cases}
\end{align*}
The entry cutoff is
\begin{equation*}
\theta_1 = \frac{c_F}{\sigma}\,\Big(\frac{c_L}{c_F}\Big)^{\frac{(1-\sigma)(\sigma+\hat{\sigma}_L)}{\hat{\sigma}_L}}.
\label{eq:theta1}
\end{equation*}
The boundary-interior switch $\theta_2$ is the unique solution to
\begin{equation*}
\MR(\theta) = \frac{c_L}{\sigma+\hat{\sigma}_L}\,
\Big(\frac{\theta\,\sigma}{c_F}\Big)^{\!\frac{\sigma(1-\sigma-\hat{\sigma}_L)}{(1-\sigma)(\sigma+\hat{\sigma}_L)}},
\label{eq:theta2-corrected}
\end{equation*}
which exists in $(1/2,1)$ if and only if $\frac{c_L}{\sigma+\hat{\sigma}_L}\,
\Big(\frac{\sigma}{c_F}\Big)^{\!\frac{\sigma(1-\sigma-\hat{\sigma}_L)}{(1-\sigma)(\sigma+\hat{\sigma}_L)}}< 1.$ Furthermore, $\theta_2>\theta_1$ if and only if $\theta_1< (\sigma+\hat{\sigma}_L)/(\sigma+2\hat{\sigma}_L)$.
Therefore, for $0<\theta_1<\theta_2<1$ to take place, the necessary and sufficient conditions are:
\begin{equation*}
\frac{c_L}{\sigma+\hat{\sigma}_L}\,
\Big(\frac{\sigma}{c_F}\Big)^{\!\frac{\sigma(1-\sigma-\hat{\sigma}_L)}{(1-\sigma)(\sigma+\hat{\sigma}_L)}}< 1,\quad \frac{c_F}{\sigma}\,\Big(\frac{c_L}{c_F}\Big)^{\frac{(1-\sigma)(\sigma+\hat{\sigma}_L)}{\hat{\sigma}_L}}<\frac{\sigma+\hat{\sigma}_L}{\sigma+2\hat{\sigma}_L}.
\end{equation*}

\noindent\emph{Efficient Allocation---}
Efficiency features single-homing with a model-switch cutoff $\hat\theta$:
\[
(q^{*}_L(\theta),q^{*}_F(\theta))=
\begin{cases}
\Big(0,\ \Big(\dfrac{\theta\,\sigma}{c_F}\Big)^{\!\frac{1}{1-\sigma}}\Big),
& \theta<\hat\theta,\\ 
\Big(\Big(\dfrac{\theta(\sigma+\hat{\sigma}_L)}{c_L}\Big)^{\!\frac{1}{1-\sigma-\hat{\sigma}_L}},\ 0\Big),
& \theta\ge \hat\theta,
\end{cases}
\]
with
\begin{equation*}
\hat\theta
=\Big(\frac{1-\sigma}{1-\sigma-\hat{\sigma}_L}\Big)^{\frac{(1-\sigma-\hat{\sigma}_L)(1-\sigma)}{\hat{\sigma}_L}}\,
\Big(\frac{\sigma}{c_F}\Big)^{\frac{\sigma(1-\sigma-\hat{\sigma}_L)}{\hat{\sigma}_L}}\,
\Big(\frac{c_L}{\sigma+\hat{\sigma}_L}\Big)^{\frac{(\sigma+\hat{\sigma}_L)(1-\sigma)}{\hat{\sigma}_L}}.
\label{eq:thetahat}
\end{equation*}

\noindent\emph{Multi-Model Monopoly---}
Virtual type is $\MR(\theta)=2\theta-1$.
Types $\theta<1/2$ are excluded. Among served types the monopolist assigns a single model, switching at $\theta^{\mathrm{m}} = (1+\hat\theta)/2$.
Optimal quantities are
\begin{align*}
(q^{\mathrm{m}}_L(\theta),q^{\mathrm{m}}_F(\theta))=
\begin{cases}
(0,0), & \theta<\tfrac12,\\ 
\Big(0,\ \Big(\dfrac{\sigma\,\MR(\theta)}{c_F}\Big)^{\!\frac{1}{1-\sigma}}\Big),
& \tfrac12\le\theta<\theta^{\mathrm{m}},\\ 
\Big(\Big(\dfrac{(\sigma+\hat{\sigma}_L)\,\MR(\theta)}{c_L}\Big)^{\!\frac{1}{1-\sigma-\hat{\sigma}_L}},\ 0\Big),
& \theta\ge\theta^{\mathrm{m}}.
\end{cases}
\end{align*}

\newpage

\newpage
\section{Contractible Tasks}\label{app:alloc}

Throughout the manuscript, we focus on the more realistic case in which the LLM provider cannot contract on the allocation of tokens across the buyer's tasks. However, the full-contracting setting provides a natural benchmark and we study it in this section. Thus, the seller designs a direct menu that specifies a token allocation across different tasks:
\begin{equation}\label{eq:menu_flexible}
    \{((x_{i1}(\w),\dots,x_{iJ}(\w))_{i\in [0,1]},z_1(\w),\dots,z_K(\w), t(\w))\}_{\w}.
\end{equation}

The allocation being contractible means the buyer has no freedom to reallocate tokens across tasks. This naturally increases the scope for screening. This also makes the problem intractable in full generality, because the reduction to a one-dimensional aggregate type and quality as in \autoref{lem:buyer-optimal-package} does not apply. Nevertheless, in this section we obtain complete characterizations in two special settings: the case of separable type distributions (\autoref{sec:cost-based}) and the case of two types (\autoref{sec:two_types}). In the former case, the optimal solution can be obtained via a menu of token budgets. In the latter case, the optimal solution can be obtained by identifying the structure of binding incentive constraints.

\subsection{Separable Type Distribution and Cost-Based Pricing} \label{sec:cost-based}
In this section, we provide a sufficient condition on type distribution under which contracting on tasks is not profitable. In fact, under that condition an even less contractually restrictive class of mechanisms, cost-based tariffs, is optimal:

\begin{definition}[Cost-Based Tariff]
A cost-based tariff is a menu of monetary budgets and transfers $\{(B_n,T_n)\}_n$ such that, upon purchasing item $n$, the buyer pays $T_n$ for access to budget $B_n$, which he can freely spend  on tokens priced at their marginal costs.
\end{definition}

We build on the analysis of \cite{arms96} 
and introduce the conditions that are jointly sufficient for the optimality of cost-based tariffs: demand separability and type separability. We show that the former is always satisfied in our setting, whereas the latter is equivalent to the aggregate type not being informative about the relative task weights.

To this end, consider a problem of optimal token allocation by type $\w$  given a monetary budget $B$ and token prices equal to marginal costs:
\begin{equation*}
  V(\w,B)=\max_{\{(x_i)_{i\in[0,1]},z\ge0\}}
  \int_{0}^{1} w_i g(x_i,z)di,\quad\mathrm{s.t.\ }\int_{0}^{1}\sum_{j=1}^{J} c_j x_{ij} di+\sum_{k=1}^{K}\hat c_k z_k
  =B,
\end{equation*} 

\begin{definition}[Demand Separability]
Demand is separable if there exist functions $V_1(\w)$ and $V_2(B)$ such that
\begin{equation}\label{eq:demand-separability}
    V(\w,B)= V_1(\w)V_2(B).
\end{equation}
\end{definition}
If demand is separable, then faced with a cost-based tariff,  all types $\w$ with the same $V_1(\w)$ purchase the same item. 

Demand separability always holds in our setting. Indeed, by
\autoref{lem:buyer-optimal-package}, if type $\w$ purchases a total amount of inference tokens $X=(X_1,\dots,X_J)$ and fine-tuning tokens $Z=(Z_1,\dots,Z_K)$, then his optimal payoff is $\theta(\w)g(X,Z)$. Therefore, (\ref{eq:demand-separability}) holds with $V_1(\w)=\theta(\w)$ and 
\begin{equation*}
V_2(B)
=\max_{X\ge 0,Z\ge 0}\ g(X,Z),\quad\mathrm{s.t.}\ \sum_{j=1}^J c_j X_j+\sum_{k=1}^K \hat c_k Z_k=B.
\end{equation*}

\begin{definition}[Type Separability]\label{ass:separability_2} Type distribution is separable if
$f_1, f_2$ exist such that $f(w)=f_1(\theta(\w))f_2(w)$ for all $w$ and $f_2$ is homogeneous of degree zero.
\end{definition}
A separable type distribution means that knowing $\theta(\w)$ provides no information about which ray from the origin $w$ lies on, that is, about the relative weights the buyer assigns to different tasks. Equivalently, denoting by $\|\cdot\|_p$  a standard $L_p$ norm, $\theta(\w)=\|w\|_{1/(1-\sigma)}$ and the type separability is equivalent to  $\|w\|_{1/(1-\sigma)}$ and $w/\|w\|_{1/(1-\sigma)}$ to be independent. Thus, any type distribution generated by a draw of a ``total size'' $\|w\|_{1/(1-\sigma)}$ according to an arbitrary distribution together with an independent draw of ``relative weights'' $w/\|w\|_{1/(1-\sigma)}$ according to any other distribution, is separable.

\begin{proposition}[Cost-Based Optimality]\label{prop:cost-based}
If the type distribution is separable, then a cost-based tariff is optimal. 
\end{proposition}

\begin{proof}
Since in our setting demand is always separable, the result follows from the arguments presented by \cite{arms96}. Specifically, we can relax the IC constraints between types located on different rays from the origin. Then, applying the Envelope Theorem ray by ray, we can find a solution to the relaxed problem as follows:
\begin{align*}
    (x^*,z^*)\in\arg\max_{\{(x_i)_{i\in[0,1]},z\ge0\}} \Bigl(1-\frac{\int_1^\infty \ind^{n-1}f(\ind w)d\ind}{f(w)}\Bigr)u(w,x,z)-c(x,z).
\end{align*}
Given the demand separability, this condition can be rewritten as:
\begin{align*}
    (x^*,z^*)\in\arg\max_{\{(x_i)_{i\in[0,1]},z\ge0\}} u(w,x,z),\quad \mathrm{s.t.\ } c(x,z)\leq B^*(w),
\end{align*}
and
\begin{align*}
    B^*(w)\in\arg\max_{B\geq 0} \Bigl(1-\frac{\int_1^\infty \ind^{n-1}f(\ind w)d\ind}{f(w)}\Bigr)\theta(\w) V_2(B)-B.
\end{align*}
At the same time, the optimal cost-based tariff implements an allocation
\begin{align*}
    (x^*,z^*)\in\arg\max_{\{(x_i)_{i\in[0,1]},z\ge0\}}  u(w,x,z),\quad\mathrm{s.t.\ }c(x,z)\leq B^*(w),
\end{align*}
where
\begin{align*}
    B^*(w)\in\arg\max_B \Bigl(\theta(\w)-\frac{1-F(\theta(\w))}{f(\theta(\w))}\Bigr) V_2(B)-B.
\end{align*}
If the type distribution is separable, then
\begin{align*}
    \Bigl(1-\frac{\int_1^\infty \ind^{n-1}f(\ind w)d\ind}{f(w)}\Bigr)\theta(\w)=\theta(\w)-\frac{1-F(\theta(\w))}{f(\theta(\w))},
\end{align*}
and the cost-based tariff implements the solution to the relaxed problem. Therefore, it is (indirectly) optimal. \end{proof}

\subsection{Binary Types}\label{sec:two_types}
Alternatively, let there be only two types, $\wone$ and $\wtwo$, which occur with prior strictly positive probabilities $f_1$ and $f_2$, respectively. In this case, the optimal mechanism depends on what happens if the seller attempts to extract the first-best level of surplus, that is, if she offers a menu containing the efficient amounts of tokens for each type with prices equal to their respective added values. If this menu is incentive compatible, then it is clearly optimal, and we call the type with the higher payment ``high'' and associate the label $H$ with it. If this menu is not incentive compatible, then we call ``high'' the type whose incentive constraint is violated. We call the other type ``low'' and associate the label $L$ with it.

To design an optimal menu, it is important to determine which type out of $\wone$ and $\wtwo$ is high and which one is low. If $\wtwo$ dominates $\wone$ for every task, then it is clearly high; alternatively, the types are in some sense horizontally differentiated and the distinction is less clear. However, it turns out that the ranking can be derived from the aggregate types, and the optimal menu admits a simple characterization.\newpage

\begin{proposition}[Binary Types]\label{prop:two_types}
The high type is the one with the higher aggregate type, i.e., $\theta(\wH)\geq\theta(\wL)$.       In the optimal menu, the token allocation of $\wH$ is always efficient. If
\begin{equation}\label{eq:two-types-cond}
    \int_0^1  (w_{Hi}-w_{Li})  w_{Li}^{\frac{\sigma}{1-\sigma}}di \leq 0,
\end{equation}
then the token allocation of $\wL$ is also efficient and the seller extracts full surplus. Otherwise, the token allocation of $\wL$ is efficient with respect to a virtual type $\wL-(\wH-\wL)f_H/f_L$.
\end{proposition}
\begin{proof}
With a small abuse of notation relative to previous sections, denote by $q=(q_i)_{i\in[0,1]}$ the \emph{profile} of qualities delivered to each task, $q_i\triangleq g(x_i,z)$, and denote by $C(q)$ the minimal total cost of generating a given profile $q$:
\begin{equation*}
     C(q)\triangleq \min_{x_i,z\geq0} \int_{0}^{1}\sum_{j=1}^{J} c_j x_{ij} di+\sum_{k=1}^{K}\hat c_k z_k,\quad\mathrm{s.t. } \ g(x_i,z) =q_i,  \forall\,i\in[0,1].
\end{equation*}

Because the set of feasible profiles $q$ is convex, it follows from the analysis of \cite*{hasi24} on general screening problems with two buyer types that in our setting either (i) the seller extracts full surplus; or (ii) the incentive constraint of type $\wH$ and the individual rationality constraint of type $\wL$ bind.

It follows that if the seller cannot extract full surplus, then 
the seller's problem can be  written as 
\begin{align*}
     \max_{q_L,q_H,t_L,t_H}& f_L(t_L-C(q_L))+f_H(t_H-C(q_H))\\
    \mathrm{s.t.}& \int_0^1 w_{Hi} q_{Hi} di -t_H=\int_0^1 w_{Hi} q_{Li} di -t_L,\  \int_0^1 w_{Li} q_{Li} di -t_L=0.
\end{align*}
Solving for transfers from the constraints, the problem can be restated as:
\begin{equation*}
     \max_{q_L,q_H} f_L \lb \int_0^1 \lb w_{Li}-\frac{f_H}{f_L} (w_{Hi}-w_{Li}) \rb q_{Li} di - C(q_L) \rb +f_H \lb \int_0^1 w_{Hi} q_{Hi} di - C(q_H) \rb,
\end{equation*}
which is solved by the allocation efficient relative to the virtual types.

It is left to determine which type is high and provide conditions for full surplus extraction. Consider a mechanism that attempts  full surplus extraction. By the efficiency analysis behind \autoref{prop:efficient},
 under this mechanism type $\w$, when reporting type $\tilde{\w}$, obtains added value
\begin{equation*}
 u(\w,\tilde{\w})=\int_{0}^{1} w_i \Psi(x_i(\tilde w)) \Phi(z(\tilde w)) di=\int_0^1 w_i \tilde w_i^{\frac{\sigma}{1-\sigma}} \Psi(d) \Phi(z(\tilde{\w}))^{\frac{1}{1-\sigma}} di.
\end{equation*}
Because under truth-telling each type obtains zero rents, the corresponding payment is $t(\w)=u(\w,\w)$, and the incentive constraint $ u(\w,\w)-t(\w)\geq u(\w,\tilde{\w})-t(\tilde\w)$ is violated if and only if:
\begin{equation}\label{eq:ic-eff}
    \Psi(d) \Phi(z(\tilde{\w}))^{\frac{1}{1-\sigma}}\int_0^1  (w_i-\tilde w_i) \tilde w_i^{\frac{\sigma}{1-\sigma}}di > 0.
\end{equation}
By Jensen's inequality and the concavity of the logarithm:
\begin{equation*}
     w_i \tilde w_i^{\frac{\sigma}{1-\sigma}}=w_i^{\frac{1-\sigma}{1-\sigma}}\tilde w_i^{\frac{\sigma}{1-\sigma}}  \leq (1-\sigma) w_i^{\frac{1}{1-\sigma}}+\sigma \tilde w_i^{\frac{1}{1-\sigma}}.
\end{equation*}
Therefore, inequality (\ref{eq:ic-eff}) implies
\begin{equation*} 
    \int_0^1  (w_i^{\frac{1}{1-\sigma}}- \tilde  w_i^{\frac{1}{1-\sigma}})di > 0,
\end{equation*}
and the incentive violation is possible only if $\theta(\w)>\theta(\tilde \w)$, i.e., from high to low aggregate type. Therefore, if full surplus extraction is not incentive compatible, then the high type is the one with the higher aggregate type. If full surplus extraction is incentive compatible, then the high type is the one with the higher aggregate type directly by the analysis behind \autoref{prop:efficient}.  The result follows.
\end{proof}

Note that even though the virtual types in \autoref{prop:two_types} follow the standard formula of the single-dimensional case, each type there is infinite-dimensional.    

One might wonder whether the correspondence between the incentive order and aggregate types extends beyond the binary-type case. This is not the case. In \autoref{sec:value-scale}, we study a special case of our model in which the buyer's type can be effectively parameterized by two variables: the number of ex ante homogeneous tasks to which the buyer attaches positive value, and the value he attaches to each of those tasks. We show that the incentive constraints that bind in the optimal mechanism do not admit a one-dimensional structure. Instead, they form an infinite collection of one-dimensional segments. In that setting the optimal mechanisms also admit a natural implementation via a menu of two-part tariffs. However, in contrast to the indirect implementations described in \autoref{sec:two-part-tariff}, each item in the menu must be accompanied by \emph{task caps}, that is, restrictions on the number of tasks the buyer can process.

\subsection{Value-Scale Heterogeneity}\label{sec:value-scale}
In this section, we characterize an optimal menu with contractible token allocations across tasks in the case of value-scale heterogeneity. In this case, each multidimensional type $\w$ is characterized (with a small abuse of notation) by two parameters $(w,s)$ such that
\begin{align}
    w_i=
    \begin{cases}
        w,\ \mathrm{if\ }i\leq s,\\
        0,\ \mathrm{if\ }i>s,
    \end{cases}
\end{align}
in which $w$ and $s$ are independently distributed according to CDFs $F_w$ and $F_s$, with $F_w$ featuring increasing virtual values. We will argue that in this case, under additional assumptions, the binding incentive constraints are those within each scale, and the seller is able to generate the same profit as if the scale were observable. 

For this section, we drop the product structure and the homogeneity requirements of (\ref{eq:gain-function}) and instead require that  $g(x_i,z)$ is  (i) positive and continuous on $\reals_{+}^{J+K}$, (ii)  strictly monotone,  twice continuously differentiable, strictly concave with negative definite Hessian, and with all cross-partial derivatives strictly positive on $\reals_{++}^{J+K}$, (iii) $g(0)=0$, 
 and (iv) Inada at zero, i.e., $\lim_{y_{m}\downarrow 0}g_{y_{m}} (y_m,y_{-m})=+\infty$ for all $y_{-m}\in\reals_+^{J+K-1}$ such that $g(0,y_{-m})>0$.\footnote{The Inada condition is imposed only for simplicity of arguments in \autoref{lem:submodular_costs}.}

Consider  the problem in which the scale is commonly known to be $s>0$. The buyer's payoff from any given item on the menu (\ref{eq:menu_flexible}) is 
\begin{equation*}
    w\, \int_{i=0}^s  g(x_i,z)di-t.
\end{equation*}
For any reported $w$ the seller should optimize token allocation to deliver a promised level of (total) quality $q$,
\begin{equation*}    wq-t,
\end{equation*}
with the minimal cost function $C(q,s)$ of delivering a given quality being:
\begin{equation}\label{eq:cost_min_contractible}
    C(q,s)=\min_{(x_i)_{i\in[0,1]},z\geq0} \int_{i=0}^s\sum_{j=1}^J c_j x_{ij} di+ \sum_{k=1}^K \hat c_k z_k,\quad\mathrm{s.t. }\ \int_{i=0}^s  g(x_i,z) di=q.
\end{equation}
Since $g$ is strictly concave, the solution to this problem is achieved by allocating the inference tokens uniformly across the $s$ tasks. The   problem can  be equivalently stated as:
\begin{equation}
    C(q,s)=\min_{x,z\geq0}\ s \sum_{j=1}^J c_j x_{j} + \sum_{k=1}^K \hat c_k z_k,\quad
    \mathrm{s.t. }\ sg(x,z) =q.\label{eqn:eqcons}
\end{equation}
The resulting cost function $C(q,s)$ satisfies, over the domain of admissible $(q,s)$, the following properties:

\begin{lemma}[Cost Function]\label{lem:submodular_costs}\strut
\begin{enumerate}
    \item $C(q,s)$ is strictly increasing and strictly convex in $q$ with $C_q(0,s)=0$.
    \item $C(q,s)$ is strictly decreasing in $s$ for $q>0$.
    \item $C(q,s)$ is submodular, i.e., $C_q(q,s)$ is decreasing in $s$ for all $q$.
\end{enumerate}
\end{lemma}
\begin{proof}
The first two statements follow directly from our assumptions on $g$. To establish the third property, posit the Lagrangian for the cost minimization problem:
\begin{equation*}
    L= s \sum_{j=1}^J c_j x_{j} + \sum_{k=1}^K \hat c_k z_k+\lambda(q- s  g(x,z)).
\end{equation*}
By the Envelope Theorem, $C_q(q,s)=\lambda(q,s)$. Thus, it suffices to show that $\lambda(q,s)$ is decreasing in $s$ for a fixed $q$. 

To this end, for any $q>0$, by the Inada condition, the solution must be interior, $(x,z)\gg 0$. The constraint binds, so $\lambda>0$. Denoting $(x,z)$ by $y$, the first-order conditions are:
\begin{equation}\label{eq:submodular-1}
    g_{x_j}(y)=c_j/\lambda,\quad g_{z_k}(y)=\hat c_k/(\lambda s).
\end{equation}
Define the Hessian function $H(y)\triangleq \nabla^2 g(y)$. Denote by $A,B\in\reals^{J+K}$ vectors such that $A_\ind=c_\ind/\lambda^2$ if $\ind\leq J$ and $=\hat c_{\ind-J}/(\lambda^2 s)$ if $\ind>J$, $B_\ind=0$ if $\ind\leq J$ and $=\hat c_{\ind-J}/(\lambda s^2)$ if $\ind>J$. Then, differentiating (\ref{eq:submodular-1}) with respect to $s$, we obtain:
\begin{equation}\label{eq:submodular-2}
    H\frac{d y}{ds}=-A\frac{d\lambda}{ds}-B.
\end{equation}
At the same time, differentiating the constraint $sg(y(s))=q$ with respect to $s$, we obtain:
\begin{equation}\label{eq:submodular-3}
   g(y)+s\nabla g(y)^\top  \frac{d y}{ds}=0.
\end{equation}
Solving for $dy/ds$ in (\ref{eq:submodular-2}) and substituting it into (\ref{eq:submodular-3}), we obtain, omitting the dependence on $y$:
\begin{equation}\label{eq:submodular-4}
    \frac{d\lambda}{ds} s\nabla g^\top (-H^{-1}A)=-g-s\nabla g^\top (-H^{-1}B).
\end{equation}

Now, observe that for all $y\in\reals_{++}^{J+K}$, by assumption on $g$, $H(y)$ is symmetric negative definite; moreover, by the positivity of cross-derivative, all off-diagonal entries of $H(y)$ are strictly positive. Then, $-H(y)$ is a Stieltjes matrix and, consequently, all elements of $H(y)^{-1}$ are negative. It immediately follows from (\ref{eq:submodular-4}) that whenever $q>0$ and $s>0$, $d\lambda/ds< 0$. 
\end{proof}

As such, the seller's problem for any given $s$ is analogous to \cite{mussa1978monopoly}. Since $w$ and $s$ are independently distributed, we can drop the dependence on $s$ and define the virtual value as
\begin{equation}
\MR(w)\triangleq w-\frac{1-F_w(w)}{f_w(w)}.
\end{equation}
Since $\MR(w)$ is increasing, all $w$ with $\MR(w)\leq0$ are excluded and all other $w$ receive the quality level  $q(w,s)$ that solves:
\begin{equation}\label{eq:quality_contractible}
    \MR(w)=C_q(q(w,s),s).
\end{equation}
The corresponding optimal transfers are
\begin{equation}\label{eq:transfers_contractible}
    t(w,s)=w\,q(w,s)-\int_0^w q(\ind,s) d\ind.
\end{equation}

\autoref{lem:submodular_costs} shows that it is cheaper to generate an extra unit of (total) quality when you have more tasks. This property is intuitive given that the returns on each task are diminishing. Thus, the optimal quality, and hence the buyer's rent, increase in scale for any given value $w$.

\begin{assumption}[Bounded Rent Increase]\label{ass:upward_cond} For all $w,s$, the function  $q(w,s)$ defined in (\ref{eq:quality_contractible}) satisfies $\int_0^{w} 
sq_s(\ind,s)d\ind\leq w q(w,s)$.

\end{assumption}

 \autoref{ass:upward_cond}  requires that the buyer's rent does not grow too quickly and, specifically, that the marginal increase of buyer rent from having an additional task is smaller than the average equilibrium value generated by LLM across existing tasks.
\begin{proposition}[Optimal Menu of Token Allocations]\label{prop:optimal_contractible}
Under 
\autoref{ass:upward_cond}, an optimal menu is
$$((x_i(w,s))_{i\in [0,1]},z(w,s), t(w,s))_{(w,s)},$$
where for each $(w,s)$, $((x_i(w,s))_{i\in [0,1]},z(w,s))$ are cost-minimizing tokens from (\ref{eq:cost_min_contractible}) that deliver quality $q(w,s)$ as defined in (\ref{eq:quality_contractible}), and  $t(w,s)$ is as defined in (\ref{eq:transfers_contractible}).
\end{proposition}
\begin{proof}
If each type reports truthfully, then the menu attains the profits of the observable-scale benchmark and is thus optimal.

If type $(w,s)$ deviates to $(w,\ts)$ with $\ts\leq s$, then, under the proposed menu, he obtains exactly the  same payoff as type $(w,\ts)$, because he processes the same number of tasks with the same willingness to pay for quality. By \autoref{lem:submodular_costs}, $C_q(q,s)$ is decreasing in $s$ for all $q$, and thus $q(w,s)$ is increasing in $s$ for all $w$. Therefore, the rents accrued by type $(w,s)$ under truth-telling,
\begin{equation*}
    U(w,s)=\int_0^w q(\ind,s) d\ind,
\end{equation*}
are increasing in $s$ for all $w$. Therefore, $(w,s)$ does not want to deviate to $(w,\ts)$ with $\ts\leq s$. Furthermore, by incentive compatibility within a given $\ts$,  $(w,\ts)$ does not want to deviate to $(\tilde w,\ts)$. Therefore, $(w,s)$ does not want to deviate to any $(\tilde w,\ts)$ with $\ts\leq s$. 

If type $(w,s)$ deviates to $(\tilde w,\ts)$ with $\ts>s$, then he obtains gross payoff $w q(\tilde w,\ts)\, s/\ts$ and pays the transfer $t(\tilde w,\ts)$. Therefore, the optimal double deviation strategy for a misreporting type solves 
\begin{equation*}\max_{\tilde w\geq0}\left[ w q(\tilde w,\ts)\frac{s}{\ts}-\tilde w q(\tilde w,\ts)+\int_0^{\tilde{w}} 
 q(\ind,\ts)d\ind\right].
\end{equation*} 
 Because the mechanism incentivizes truthful reporting by any $\ts$-truthtelling type, including the type $(w s/\ts,\ts)$, it follows that 
\begin{equation*}
\tilde w^*=\frac{w s}{\ts},\quad U(w;s,\ts)= \int_0^{\frac{ws}{\ts}} 
 q(\ind,\ts)d\ind.
 \end{equation*} 
 The  condition that   discourages local deviations to $\ts>s$ is:  
 \begin{equation*}s \int_0^{w} 
 q_s(\ind,s)d\ind\leq w q(w,s),
 \end{equation*}
 for all $(w,s)$, which is precisely \autoref{ass:upward_cond}.
Furthermore, observe that for $\ts>s$:
\begin{align*}
    U_{\ts}(w;s,\ts)&=\int_0^{\frac{ws}{\ts}} 
 q_{s}(\ind,\ts)d\ind-\frac{ws}{\ts^2}q(\frac{ws}{\ts},\ts)\\
 &=\frac{1}{\ts}\left(\ts \int_0^{\frac{ws}{\ts}} 
 q_{s}(\ind,\ts)d\ind-\frac{ws}{\ts}q(\frac{ws}{\ts},\ts)\right)\leq0,
\end{align*}
where the inequality holds by \autoref{ass:upward_cond} with  $(w,s)$ replaced by $(sw/\ts,\ts)$. Therefore, the upward global deviations in $\ts$ are suboptimal and the result follows.
\end{proof}

\newpage
\bibliography{general_llm}

\begin{thebibliography}{24}
\newcommand{\enquote}[1]{``#1''}
\expandafter\ifx\csname natexlab\endcsname\relax\def\natexlab#1{#1}\fi

\bibitem[\protect\citeauthoryear{Armstrong}{Armstrong}{1996}]{arms96}
\textsc{Armstrong, M.} (1996): \enquote{Multiproduct Nonlinear Pricing,}
  \emph{Econometrica}, 64, 51--76.

\bibitem[\protect\citeauthoryear{Babaioff, Kleinberg, and Paes~Leme}{Babaioff
  et~al.}{2012}]{babaioff2012optimal}
\textsc{Babaioff, M., R.~Kleinberg, and R.~Paes~Leme} (2012): \enquote{Optimal
  Mechanisms for Selling Information,} in \emph{Proceedings of the 13th ACM
  Conference on Electronic Commerce}, 92--109.

\bibitem[\protect\citeauthoryear{Bergemann, Bojko, D{\"u}tting, Paes~Leme, Xu,
  and Zuo}{Bergemann et~al.}{2024}]{bergemann2024data}
\textsc{Bergemann, D., M.~Bojko, P.~D{\"u}tting, R.~Paes~Leme, H.~Xu, and
  S.~Zuo} (2024): \enquote{Data-Driven Mechanism Design: Jointly Eliciting
  Preferences and Information,} \emph{arXiv preprint arXiv:2412.16132}.

\bibitem[\protect\citeauthoryear{Bergemann, Bonatti, and Smolin}{Bergemann
  et~al.}{2018}]{bergemann2018design}
\textsc{Bergemann, D., A.~Bonatti, and A.~Smolin} (2018): \enquote{The Design
  and Price of Information,} \emph{American Economic Review}, 108, 1--48.

\bibitem[\protect\citeauthoryear{Calzolari and Denicol{\`o}}{Calzolari and
  Denicol{\`o}}{2013}]{cade13}
\textsc{Calzolari, G. and V.~Denicol{\`o}} (2013): \enquote{Competition with
  Exclusive Contracts and Market-Share Discounts,} \emph{American Economic
  Review}, 103, 2384--2411.

\bibitem[\protect\citeauthoryear{Calzolari and Denicol{\`o}}{Calzolari and
  Denicol{\`o}}{2015}]{cade15}
---\hspace{-.1pt}---\hspace{-.1pt}--- (2015): \enquote{Exclusive Contracts and
  Market Dominance,} \emph{American Economic Review}, 105, 3321--3351.

\bibitem[\protect\citeauthoryear{Castro-Pires, Chade, and
  Swinkels}{Castro-Pires et~al.}{2024}]{castro2024disentangling}
\textsc{Castro-Pires, H., H.~Chade, and J.~Swinkels} (2024):
  \enquote{Disentangling Moral Hazard and Adverse Selection,} \emph{American
  Economic Review}, 114, 1--37.

\bibitem[\protect\citeauthoryear{Daskalakis, Deckelbaum, and Tzamos}{Daskalakis
  et~al.}{2017}]{daskalakis2017strong}
\textsc{Daskalakis, C., A.~Deckelbaum, and C.~Tzamos} (2017): \enquote{Strong
  Duality for a Multiple-Good Monopolist,} \emph{Econometrica}, 85, 735--767.

\bibitem[\protect\citeauthoryear{Demirer, Fradkin, Tadelis, and Peng}{Demirer
  et~al.}{2025}]{defr25}
\textsc{Demirer, M., A.~Fradkin, N.~Tadelis, and S.~Peng} (2025): \enquote{The
  Emerging Market for Intelligence: Pricing, Supply, and Demand for LLMs,}
  Tech. rep., National Bureau of Economic Research.

\bibitem[\protect\citeauthoryear{Devanur, Goldner, Saxena, Schvartzman, and
  Weinberg}{Devanur et~al.}{2020}]{devanur2020optimal}
\textsc{Devanur, N.~R., K.~Goldner, R.~R. Saxena, A.~Schvartzman, and S.~M.
  Weinberg} (2020): \enquote{Optimal Mechanism Design for Single-Minded
  Agents,} in \emph{Proceedings of the 21st ACM Conference on Economics and
  Computation}, 193--256.

\bibitem[\protect\citeauthoryear{Doligalski, Dworczak, Krysta, and
  Tokarski}{Doligalski et~al.}{2025}]{dodkt25}
\textsc{Doligalski, P., P.~Dworczak, J.~Krysta, and F.~Tokarski} (2025):
  \enquote{Incentive separability,} \emph{Journal of Political Economy
  Microeconomics}, 3, 539--567.

\bibitem[\protect\citeauthoryear{Duetting, Mirrokni, Paes~Leme, Xu, and
  Zuo}{Duetting et~al.}{2024}]{duetting2024mechanism}
\textsc{Duetting, P., V.~Mirrokni, R.~Paes~Leme, H.~Xu, and S.~Zuo} (2024):
  \enquote{Mechanism design for large language models,} in \emph{Proceedings of
  the ACM on Web Conference 2024}, 144--155.

\bibitem[\protect\citeauthoryear{Fiat, Goldner, Karlin, and Koutsoupias}{Fiat
  et~al.}{2016}]{fiat2016fedex}
\textsc{Fiat, A., K.~Goldner, A.~R. Karlin, and E.~Koutsoupias} (2016):
  \enquote{The Fedex Problem,} in \emph{Proceedings of the 2016 ACM Conference
  on Economics and Computation}, 21--22.

\bibitem[\protect\citeauthoryear{Fish, Gonczarowski, and Shorrer}{Fish
  et~al.}{2024}]{fish2024algorithmic}
\textsc{Fish, S., Y.~A. Gonczarowski, and R.~I. Shorrer} (2024):
  \enquote{Algorithmic Collusion by Large Language Models,} \emph{arXiv
  preprint arXiv:2404.00806}.

\bibitem[\protect\citeauthoryear{Haghpanah and Siegel}{Haghpanah and
  Siegel}{2024}]{hasi24}
\textsc{Haghpanah, N. and R.~Siegel} (2024): \enquote{Screening Two Types,}
  Tech. rep., Penn State.

\bibitem[\protect\citeauthoryear{Jullien}{Jullien}{2000}]{jull00}
\textsc{Jullien, B.} (2000): \enquote{Participation Constraints in Adverse
  Selection Models,} \emph{Journal of Economic Theory}, 93, 1--47.

\bibitem[\protect\citeauthoryear{Kaplan, McCandlish, Henighan, Brown, Chess,
  Child, Gray, Radford, Wu, and Amodei}{Kaplan
  et~al.}{2020}]{kaplan2020scaling}
\textsc{Kaplan, J., S.~McCandlish, T.~Henighan, T.~B. Brown, B.~Chess,
  R.~Child, S.~Gray, A.~Radford, J.~Wu, and D.~Amodei} (2020): \enquote{Scaling
  Laws for Neural Language Models,} \emph{arXiv preprint arXiv:2001.08361}.

\bibitem[\protect\citeauthoryear{Laffont and Tirole}{Laffont and
  Tirole}{1990}]{lati90}
\textsc{Laffont, J.-J. and J.~Tirole} (1990): \enquote{The regulation of
  multiproduct firms: Part I: Theory,} \emph{Journal of Public Economics}, 43,
  1--36.

\bibitem[\protect\citeauthoryear{Mahmood}{Mahmood}{2024}]{mahmood2024pricing}
\textsc{Mahmood, R.} (2024): \enquote{Pricing and Competition for Generative
  AI,} \emph{arXiv preprint arXiv:2411.02661}.

\bibitem[\protect\citeauthoryear{Mussa and Rosen}{Mussa and
  Rosen}{1978}]{mussa1978monopoly}
\textsc{Mussa, M. and S.~Rosen} (1978): \enquote{Monopoly and Product Quality,}
  \emph{Journal of Economic Theory}, 18, 301--317.

\bibitem[\protect\citeauthoryear{Rochet and Stole}{Rochet and
  Stole}{2003}]{rochet2003economics}
\textsc{Rochet, J.-C. and L.~A. Stole} (2003): \enquote{The Economics of
  Multidimensional Screening,} \emph{Econometric Society Monographs}, 35,
  150--197.

\bibitem[\protect\citeauthoryear{Tirole}{Tirole}{1988}]{tiro88}
\textsc{Tirole, J.} (1988): \emph{The Theory of Industrial Organization},
  Cambridge: MIT Press.

\bibitem[\protect\citeauthoryear{Wu, Sun, Li, Welleck, and Yang}{Wu
  et~al.}{2025}]{wu2025inference}
\textsc{Wu, Y., Z.~Sun, S.~Li, S.~Welleck, and Y.~Yang} (2025):
  \enquote{Inference Scaling Laws: An Empirical analysis of Compute-Optimal
  Inference for LLM Problem-Solving,} in \emph{The Thirteenth International
  Conference on Learning Representations}.

\bibitem[\protect\citeauthoryear{Yang}{Yang}{2022}]{yang2022selling}
\textsc{Yang, K.~H.} (2022): \enquote{Selling Consumer Data for Profit: Optimal
  Market-Segmentation Design and Its Consequences,} \emph{American Economic
  Review}, 112, 1364--1393.

\end{thebibliography}
\bibliographystyle{ecta}
\end{document}